\documentclass[aps,twocolumn,showpacs,preprintnumbers,amssymb]{revtex4}

\newcommand{\be}{\begin{eqnarray}}
\newcommand{\ee}{\end{eqnarray}}

\newcommand{\eins}{\mbox{$1 \hspace{-1.0mm}  {\bf l}$}}

\newcommand{\Bra}[1]{\mbox{$\langle #1 |$}}
\newcommand{\Ket}[1]{\mbox{$| #1 \rangle$}}
\newcommand{\ket}[1]{\mbox{$| #1 \rangle$}}

\def\bea{\begin{eqnarray}}
\def\eea{\end{eqnarray}}
\def\C{\hbox{$\mit I$\kern-.7em$\mit C$}}
\def\N{\hbox{$\mit I$\kern-.3em$\mit N$}}

\def\tr{{\rm tr}}

\usepackage{dcolumn} 
\usepackage{bm} 

\usepackage{epsfig}

\begin{document}

\title{Multiparticle entanglement purification for two-colorable graph states}

\author{H. Aschauer$^1$, W. D\"{u}r$^{2}$  and H.-J. Briegel$^{1,2,3}$}

\affiliation{
$^1$Sektion Physik, Ludwig-Maximilians-Universit\"at M\"unchen, Theresienstr.\ 37, D-80333 M\"unchen, Germany.\\ 
$^2$Institut f\"ur Theoretische Physik, Universit\"at Innsbruck, A-6020 Innsbruck, Austria.\\
$^3${Institut f\"ur Quantenoptik und Quanteninformation der \"Osterreichischen Akademie der Wissenschaften, Innsbruck, Austria.}} 

\date{\today}

\begin{abstract}
We investigate multiparticle entanglement purification schemes which allow one to purify all two colorable graph states, a class of states which includes e.g. cluster states, GHZ states and codewords of various error correction codes. The schemes include both recurrence protocols and hashing protocols. We analyze these schemes under realistic conditions and observe for a generic error model that the threshold value for imperfect local operations depends on the structure of the corresponding interaction graph, but is otherwise independent of the number of parties. The qualitative behavior can be understood from an analytically solvable model which deals only with a restricted class of errors. 
We compare direct multiparticle entanglement purification protocols with schemes based on bipartite entanglement purification and show that the direct multiparticle entanglement purification is more efficient and the achievable fidelity of the purified states is larger. We also show that the purification protocol allows one to produce private entanglement, an important aspect when using the produced entangled states for secure applications. Finally we discuss an experimental realization of a multiparty purification protocol in optical lattices which is issued to improve the fidelity of cluster states created in such systems.  
\end{abstract}


\maketitle


\section{Introduction}

In recent years a number of surprising, unexpected applications of entangled states have been developed. In the bipartite case, teleportation \cite{Be93}, superdense coding \cite{Be91} and entanglement based quantum cryptography \cite{Ek91} are well-known examples. In the multipartite case it was shown that multiparticle entangled states (MES) allow one not only to accomplish several tasks in multi--party communication scenarios ---such as secret sharing or secure function evaluation \cite{secure}--- but also to improve the precision of frequency measurements, 
leading to higher frequency standards \cite{Wi93,Hu97}. Furthermore, many error correction codes are based on MES, and certain MES ---the so--called cluster states \cite{Br01}--- have even been shown to constitute a universal resource for quantum computation when assisted by local measurements only \cite{Ra98}.

All these applications require the use of certain bipartite or multipartite entangled pure states. In reality, however, those states will not be available with unit fidelity. On the one hand, the operations required to create the states will be noisy.  On the other hand, the MES interact with the environment and will be subjected to decoherence, or the particles constituting the entangled state have to be sent through noisy quantum channels in a communication scenario with distant parties. Thus in practice only mixed states rather than pure states are available and it is a central problem to establish methods to increase the quality of the states by some means.

In principle, entanglement purification provides a method to accomplish this task. Efficient protocols to obtain a few high--fidelity entangled states from several low--fidelity entangled states by using local operations and classical communication are known. Most purification protocols for qubits introduced so far are only capable to purify a specific type of states, namely states which are equivalent up to local unitary operations to states of the form $|0\rangle^{\otimes N} + |1\rangle^{\otimes N}$, the so called ``GHZ'' states \cite{Be96,De96,Mu99,Sm00}. Only quite recently, we have introduced \cite{Du03} multiparticle entanglement purification protocols (MEPPs) which are capable of purifying all two colorable graph states, a class of multi--qubit states which will be defined below and which includes, for instance, GHZ states, cluster states and codewords of error correction codes. In this paper, we provide a detailed analysis of these protocols and provide addition material, including a hashing protocol for this class of states and a comparison of multiparticle entanglement purification with protocols based on bipartite entanglement purification.

The paper is organized as follows. In Sec. \ref{propertiesgen}, we review the concept of graph states, fix some notation and highlight a number of useful properties of these states. Sec. \ref{MEPP} is concerned with multiparticle entanglement purification protocols. On the one hand, we review the recurrence protocol introduced in Ref. \cite{Du03} in Sec. \ref{recurrence} and analyze in detail its properties. We investigate the purification regime, the convergence, as well as the efficiency of the procedure. We provide both analytic analysis for certain low rank states and a numerical analysis for generic states. On the other hand, we introduce in Sec. \ref{hashing} a hashing protocol which is capable of purifying two colorable graph states with a finite yield. In Sec. \ref{ILO} we analyze numerically the recurrence protocol for different target states ---in particular cluster states and GHZ states--- under realistic conditions using a generic error model of local control operations. We determine the purification regime, i.e. the minimum required and maximal reachable fidelity, as well as the threshold value for noise in local operations below which the purification protocol can be successfully applied. An analytic treatment for a restricted error model is carried out in Sec. \ref{binarymix}, recovering essentially the same behavior as for the generic error model. In Sec. \ref{comparebipartite}, multiparticle entanglement purification protocols are compared with protocols based on bipartite entanglement purification. We analyze both the case of noiseless local operations as well as noisy local operations. We find in the former case that direct multiparticle entanglement purification is more efficient than any scheme based on bipartite purification. In the latter case, the reachable fidelity is higher. In Sec. \ref{private} we are concerned with security aspects of our protocols and show that the purified entanglement is private. Sec. \ref{applications} deals with a number of possible applications of the purification protocols. A possible experimental implementation based on neutral atoms in an optical lattice is discussed in Sec. \ref{experiment}. We summarize and conclude in Sec. \ref{summary}.


\section{Graph states: basic principles and properties}\label{propertiesgen}

\subsection{Graph states}\label{graphstates}

In this section, we review the concept of graph states, describe some of their properties and fix notation. Graph states have first been introduced in Ref. \cite{Ra03}, generalizing the notion of cluster states as introduced in Ref. \cite{Br01}. A detailed investigation of their entanglement properties has recently been given in the paper by Hein et al. \cite{He03}. Graph states occur in various contexts in quantum information theory, in which multi-party quantum correlations play a central role. Examples are multi-party quantum communication, measurement-based quantum computation, and quantum error correction. Using terminology of standard quantum mechanics textbooks, a graph state can be described as the common eigenstate of a complete set of commuting observables . The graph associated with a given graph state can thus be regarded as a shorthand notation for its complete set of commuting observables. In quantum error correction, the set of commuting observables is also referred to as the stabilizer group of the state. Note, however, that for the purpose of quantum error correction, the stabilizer is usually not complete, since degenerate subspaces are used as code spaces. The graph codes introduced in Ref. \cite{Sc01} take account of this fact, and can be regarded as an application of graph states in the specific context of quantum error correction.

Consider a graph ${\cal G}=(V,E)$ which is a set of vertices $V$ connected in a specific way by edges $E$. The edges specify a neighborhood relation between vertices. Associated with any graph ${\cal G}$ are a set of $N=|V|$ commuting correlation operators
\be
K_j= \sigma_x^{(j)} \prod_{\{k,j\} \in E} \sigma_z^{(k)}.
\label{K}
\ee
That is, to any vertex $j$ corresponds a correlation operator $K_j$ which is given by the spin 1/2 Pauli operator $\sigma_x$ on vertex $j$, $\sigma_z$ on all neighboring vertices of $j$, i.e. all vertices $k$ which are connected to $j$  by edges, and the identity operator on the remaining vertices. Graph states associated with ${\cal G}$, $|\Psi_{\mu_1\mu_2\ldots \mu_N}\rangle$, $\mu_j \in \{0,1\}$, are the joint eigenstates of the set of hermitean operators $\{K_j| j \in V\}$ which fulfill the eigenvalue equations 
\be
K_j |\Psi_{\mu_1\mu_2\ldots \mu_N}\rangle_{\cal G} = (-1)^{\mu_j}|\Psi_{\mu_1\mu_2\ldots \mu_N}\rangle_{\cal G} ~\forall j.\label{EV}
\ee 
For notational convenience we will omit the index ${\cal G}$ whenever there is no danger of confusion, $|\Psi_{\mu_1\mu_2\ldots \mu_N}\rangle_{\cal G} \equiv |\Psi_{\mu_1\mu_2\ldots \mu_N}\rangle$.  Note that the graph states $\{|\Psi_{\mu_1\mu_2\ldots \mu_N}\rangle_{\cal G}\}$ are uniquely defined by the eigenvalue equations and form a basis in ${\cal H}=(\C^2)^{\otimes N}$, i.e. 
\bea
&&|\langle\Psi_{\mu_1\mu_2\ldots\mu_N} | \Psi_{\nu_1\nu_2\ldots\nu_N}\rangle|^2=\delta_{\mu_1\nu_1} \delta_{\mu_2\nu_2}\ldots \delta_{\mu_N\nu_N},\nonumber\\
&&\sum_{\mu_1,\mu_2,\ldots,\mu_N=0}^1 |\Psi_{\mu_1\mu_2\ldots\mu_N}\rangle \langle\Psi_{\mu_1\mu_2\ldots\mu_N}| = \eins
\eea
We remark that apart from the description of graph states by a set of commuting correlation observables, one can also give an equivalent description of the state in terms of an ``interaction graph'' \cite{Br01,Ra03}. To be specific, consider the interaction Hamiltonian 
\be
H_{kl}= (\eins^{(k)}-\sigma_z^{(k)})/2\otimes(\eins^{(l)}-\sigma_z^{(l)})/2,
\ee
which acts on particles $k$ and $l$ and corresponds, up to local unitary operations, to an Ising interaction. We consider the initial state $|\psi\rangle$ where all particles are prepared in the state $|+\rangle$ with $|+\rangle=1/\sqrt{2}(|0\rangle+|1\rangle)$, i.e. $|\psi\rangle=|+\rangle^{\otimes N}$. For a fixed graph ${\cal G}$, the corresponding graph state $|\Psi_{00\ldots 0}\rangle$ is obtained by applying on the state $|\psi\rangle$ the interaction Hamiltonian $H_{kl}$ for time $t=\pi$ to all those pairs of particles whose vertices in the corresponding graph are connected by edges, that is
\be
|\Psi_{00\ldots 0}\rangle =\prod_{(k,l)\in E} e^{-i \pi H_{kl}}|+\rangle^{\otimes N}.
\ee

Note that graph states constitute a large family of multiparticle entangled states with various entanglement properties. To be specific, for a fixed $N$ we have $2^{N(N-1)/2}$ different graphs, although not all of them are inequivalent and correspond to different kinds of entanglement (see Sec. \ref{localequivalence}). Throughout the paper, we will mainly consider two--colorable graphs, that are graphs for which a partition of the vertices into two disjoint sets $V_A \cup V_B=V$ with $N_A\equiv|V_A|, N_B\equiv|V_B|$,$N=N_A+N_B$ exists such that no vertices within one set are connected by edges (equivalently, a two--coloring of the graph with respect to its vertices exist). The states arising from such two--colorable graphs, which we call two--colorable graph states (TCGS), include a number of interesting multiparticle entangled states, e.g. Greenberger--Horne--Zeilinger (GHZ) states, cluster states or codewords of certain error correction codes. We remark that it was recently shown that two--colorable graph states are in fact equivalent to the so called Calderbank-Shor-Steane (CSS) states \cite{Rai04}. That is, any state that can be written as a two--colorable graph state can also be written (up to local unitary transformations) as a CSS state and vice versa.


\subsection{Examples}\label{example}

As a first example, consider the $N$--particle GHZ state. The graph corresponding to a $N$--particle GHZ state is given by $N$ vertices $\{1,2,\ldots, N\}$ and edges $\{1,k\}$, $k\in\{2,3,\ldots,N\}$. This graph can easily seen to be two--colorable by considering the sets $V_A=\{1\}$ and $V_B=\{2,3,\ldots, N\}$. The corresponding two--colorable graph state $|\Psi_{00\ldots 0}\rangle$ is given by
\be
|\Psi_{00\ldots 0}\rangle=1/\sqrt{2}(|0_z\rangle\otimes|0_x\rangle^{\otimes N-1}+|1_z\rangle\otimes|1_x\rangle^{\otimes N-1}),
\ee 
where $\{|0_z\rangle,|1_z\rangle\}$ [$\{|0_x\rangle,|1_x\rangle\}$] is the eigenbasis of $\sigma_z$ [$\sigma_x$] respectively, with $|0_x\rangle=1/\sqrt{2}(|0_z\rangle+|1_z\rangle)$. 

The graph corresponding to an (open) linear cluster state of length $N$ is given by $N$ vertices $\{1,2,\ldots, N\}$ and edges $\{k,k+1\}$, $k\in\{1,2,\ldots,N-1\}$, i.e. all neighboring vertices are connected by edges. In this case, the sets $V_A$ [$V_B$] are given by all odd [even] vertices respectively, which shows that the graph is two--colorable. The corresponding two--colorable graph states for arbitrary $N$ are rather difficult to write down explicitly, as the minimum number of terms required to specify the state in any product basis grows exponentially with $N$ \cite{Ra03}. This is reflected by the fact that the amount of entanglement of these states, as quantified by the Schmidt measure \cite{BrEi00}, grows linearly with $N$. For our present purposes an explicit expansion is not required, since the description in terms of the correlation operators (Eq. (\ref{EV})) is complete and all calculations can be performed using the corresponding eigenvalue equations. This is one of the main advantages of the (abstract) definition of graph states as eigenstates of a set of commuting correlation operators and it allows for a simplified analytical and numerical treatment of protocols operating on graph states. This parallels the treatment of quantum error correcting codes in terms of the stabilizer formalism \cite{Sc01}.

\begin{figure}[ht]
\begin{picture}(230,90)
\put(-5,-5){\epsfxsize=230pt\epsffile[74 717 378 821]{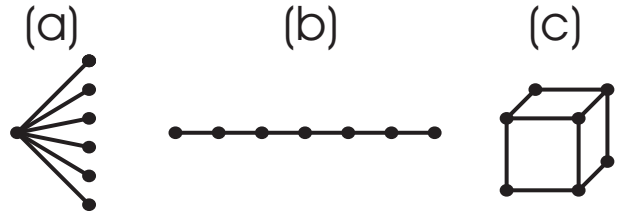}}
\end{picture}
\caption[]{Graphs with $N=7$ corresponding to (a) GHZ state, (b) linear cluster state and (c) seven qubit steane code.}
\label{cube}
\end{figure}

As a final example, consider a graph which consists of  seven vertices of a cube. The graph states associated with such a graph are equivalent, up to local unitaries, to the codewords of the seven--qubit Steane code ($[7,1,3]$ CSS code). The graphs associated with these examples are illustrated in Fig. \ref{cube}.

\subsection{Useful properties of graph states}\label{properties}

For any fixed graph ${\cal G}$ one can verify a number of useful relations between graph states following from Eq.(\ref{EV}). For any vertex $j$ we divide the vertices into three distinct sets: vertex $j$, the set $N_j$ which contains all neighboring vertices of $j$, i.e. all vertices connected to $j$, $N_j=\{k \in V | \{k,j\} \in E\}$, and the set $R_j$ which contains the remaining vertices.  We use the corresponding multindices ${\bm \mu_{N_j}}$, ${\bm \mu_{R_j}}$ and the index $\mu_j$ to label the graph states, where 
${\bm \mu_{N_j}}\equiv\mu_{k_1}\mu_{k_2}\ldots \mu_{k_{|N_j|}}$, 
${\bm \mu_{R_j}}=\mu_{i_1}\mu_{i_2}\ldots \mu_{i_{|R_j|}}$ with $\{k_l,j\} \in E$, $\{i_l,j\} \not\in E$. 
One readily verifies that for each $j$ the following relations are fulfilled
\bea
\sigma_z^{(j)}|\Psi_{\mu_j{\bm \mu_{N_j}}{\bm \mu_{R_j}}}\rangle &=& |\Psi_{\bar\mu_j{\bm \mu_{N_j}}{\bm \mu_{R_j}}}\rangle,\label{Sz}\\
\sigma_x^{(j)}|\Psi_{\mu_j{\bm \mu_{N_j}}{\bm \mu_{R_j}}}\rangle &=& (-1)^{\mu_j}|\Psi_{\mu_j \bar{\bm \mu}_{N_j}{\bm \mu_{R_j}}}\rangle,\label{Sx}\\
\sigma_y^{(j)}|\Psi_{\mu_j{\bm \mu_{N_j}}{\bm \mu_{R_j}}}\rangle &=& i(-1)^{\bar\mu_j}|\Psi_{\bar\mu_j \bar{\bm \mu}_{N_j}{\bm \mu_{R_j}}}\label{Sy}\rangle,
\eea
where ${\bm \bar\mu_{N_j}}=\bar\mu_{k_1}\bar\mu_{k_2}\ldots \bar\mu_{k_{|N_j|}}$ denotes the bitwise complement with $\bar 0=1, \bar 1=0$. Eq. (\ref{Sz}) implies that
\be
|\Psi_{\mu_1\mu_2\ldots \mu_N}\rangle = \sigma_z^{\mu_1}\sigma_z^{\mu_2}\ldots \sigma_z^{\mu_N}|\Psi_{00\ldots 0}\rangle, 
\ee
where $\sigma_z^0=\eins$. This property follows from the eigenvalue equations Eq. (\ref{EV}), while
Eq. (\ref{Sx}) follows from
\bea 
&&\sigma_x^{(j)}|\Psi_{\mu_j{\bm \mu_{N_j}}{\bm \mu_{R_j}}}\rangle= 
(-1)^{\mu_j}\sigma_x^{(j)}K_j|\Psi_{\mu_j{\bm \mu_{N_j}}{\bm \mu_{R_j}}}\rangle=\nonumber\\
&=&(-1)^{\mu_j}\sigma_z^{(k_1)}\sigma_z^{(k_2)}\ldots \sigma_z^{(k_{|N_j|})}|\Psi_{\mu_j{\bm \mu_{N_j}}{\bm \mu_{R_j}}}\rangle\nonumber\\
&=&(-1)^{\mu_j}|\Psi_{\mu_j\bar {\bm \mu}_{N_j}{\bm \mu_{R_j}}}\rangle.
\eea
Finally, to prove Eq. ({\ref{Sy}) one uses that $\sigma_y^{(j)}=i\sigma_x^{(j)}\sigma_z^{(j)}$ together with Eqs. (\ref{Sz},\ref{Sx}).


\subsection{Local equivalence of graph states}\label{localequivalence}

While different multiparticle entangled graph states are associated with different graphs, it is not obvious that states arising from different ``interaction'' graphs lead to states with different entanglement properties. In fact, it turns out that local unitary operations allow one to change from some graph state to some other. The classification of graph states into subclasses that are invariant under local unitary transformations is a complex problem, which is not solved in general. Progress among these lines is reported in Refs. \cite{He03,Ne03}. We emphazise that the results we obtain below for certain graph states, in particular for all two--colorable graph states, are also valid for graph states which are local unitary equivalent to these graphs. This implies that the entanglement purification protocols discussed below are applicable to some graph states which do not arise from a two--colorable graph. For instance, the GHZ--state discussed in Sec. \ref{example} associated with a graph with edges $\{1,k\} \forall k$ is local unitary equivalent to a state associated with the fully connected graph, i.e. with edges $\{k,l\}, \forall k<l$. While the first graph is clearly two--colorable, the second is not.


\subsection{Mixed states and depolarization}\label{depolarization}

Let us now consider an arbitrary graph ${\cal G}$ with $N$ vertices $V=\{V_1,V_2,\ldots,V_N\}$, and $N$ spatially distinct parties each holding one of the $N$ particles belonging to a general mixed state $\rho$. We consider the $N$--particle graph states $\{|\Psi_{\mu_1\mu_2\ldots\mu_N}\rangle_{\cal G}\}$ associated to ${\cal G}$ and introduce the multi--index ${\bm \mu} \equiv \mu_1\mu_2\ldots \mu_N$.  Since these states form a basis in $\cal H$ the density operator $\rho$ can be expressed as
\be
\rho=\sum_{{\bm \mu},{\bm \nu}} \lambda_{{\bm \mu},{\bm \nu}}|\Psi_{{\bm \mu}}\rangle_{\cal G}\langle\Psi_{{\bm \nu}}|.\label{rhonat}
\ee
In the following, we will show that one can depolarize $\rho$ to a state $\rho_{\cal G}$ which is diagonal in the graph state basis by a sequence of local operations and classical communication (i.e. operations acting on each particle individually), without changing the diagonal coefficients. That is, given $\rho$ (Eq. (\ref{rhonat})) one can create by means of local operations and classical communication the state 
\be
\rho_{\cal G}=\sum_{\bm \mu} \lambda_{{\bm \mu}}|\Psi_{{\bm \mu}}\rangle\langle\Psi_{{\bm \mu}}|
\ee
with $\lambda_{{\bm \mu}}\equiv \lambda_{{\bm \mu},{\bm \mu}}$.

This can easily be seen using the eigenvalue equation Eq. (\ref{EV}). Consider two graph states $|\Psi_{\mu_1\mu_2\ldots \mu_N}\rangle$ and $|\Psi_{\nu_1\nu_2\ldots \nu_N}\rangle$ which differ in at least one bit, say the first $\mu_1=0$ while $\nu_1=1$. We have that $K_1 |\Psi_{0\mu_2\ldots \mu_N}\rangle = (+1)  |\Psi_{0\mu_2\ldots \mu_N}\rangle$ and $K_1 |\Psi_{1\nu_2\ldots \nu_N}\rangle = (-1) |\Psi_{1\nu_2\ldots \nu_N}\rangle$. Note that the operation corresponding to $K_1$ is local, i.e. involves only operations on individual particles. If the parties thus jointly perform with probability $p=1/2$ the operations corresponding to $K_1$, while with probability $p=1/2$ the state is left untouched, the resulting density operator $\tilde \rho =1/2(\rho+ K_1\rho K_1^\dagger)$ will have no off diagonal elements of the form $|\Psi_{0\mu_2\ldots \mu_N}\rangle\langle \Psi_{1\nu_2\ldots \nu_N}|$, while the diagonal elements remain unchanged. In a similar way, all off diagonal elements can be eliminated in a total of $N$ rounds by probabilistically applying the local operations corresponding to $K_j, j=1,2,\ldots ,N$ to the state resulting from the previous round.

In summary, for any graph one can depolarize the state $\rho$ to a mixed state $\rho_{\cal G}$ diagonal in the associated graph state basis. The corresponding sequence of (probabilistic) local operations is determined by the correlation operators $K_j$ associated with the graph ${\cal G}$. This ensures that we can restrict ourselves to mixed states diagonal in the graph state basis in the following analysis.


\section{Multiparticle entanglement purification protocols}\label{MEPP}

In the following, we will analyze in detail the multiparticle entanglement purification protocol introduced in Ref. \cite{Du03}. This protocol is a recurrence--like scheme which operates on two copies of a given state simultaneously and may be viewed as a generalization of the purification protocol 
for GHZ states introduced in Ref. \cite{Mu99} to arbitrary two-colorable graph states. We will also introduce a multi--party hashing protocol ---based on the protocol presented in Ref. \cite{Sm00} for GHZ states---, where joint manipulations of a large number of copies are involved. In both cases, the goal is to produce few states with high fidelity from a large number of states with low fidelity. While the first protocol is particularly useful to purify states of low fidelity, the second protocol turns out to be very efficient for states sufficiently close to the desired output state. We investigate the conditions under which the protocols can be applied and also discuss their efficiencies.

In the following, we consider an arbitrary but fixed two--colorable graph ${\cal G}$ with vertices $V=V_A\cup V_B$, $N_A\equiv|V_A|, N_B\equiv|V_B|$ and $N=N_A+N_B$ spatially distinct parties each holding one of the $N$ particles that belong to a general mixed state $\rho$. Using the depolarization procedure discussed in the previous section, we can transform the state $\rho$ to a standard form diagonal in the associated graph state basis, without changing the diagonal coefficients. That is, without loss of generality, we can consider mixed states $\rho$ diagonal in the graph--state basis
\be
\rho=\sum_{{\bm \mu}_{\bf A},{\bm \mu}_{\bf B}} \lambda_{{\bm \mu}_{\bf A},{\bm \mu}_{\bf B}} |\Psi_{{\bm \mu}_{\bf A},{\bm \mu}_{\bf B}}\rangle\langle \Psi_{{\bm \mu}_{\bf A},{\bm \mu}_{\bf B}}|.
\ee
We have introduced the shorthand notation ${{\bm \mu}_{\bf A}} \equiv \mu_{i_1}\mu_{i_2} \ldots \mu_{i_{N_{A}}}$ for all eigenvalues associated with the vertices in the set $V_A$, and similar for ${{\bm \mu}_{\bf B}}$. We assume that the parties share $M$ copies of this $N$--particle mixed state $\rho$. In the following we establish for every two--colorable graph ${\cal G}$ a local purification protocol which is capable of creating the pure state $|\Psi_{{\bf 0}}\rangle_{\cal G}$ as output state, given the initial state $\rho$ fulfills certain requirements (e.g. has sufficiently high fidelity). Note that we have used the shorthand notation ${\bf 0}\equiv 00\ldots 0$, i.e. $|\Psi_{{\bf 0}}\rangle_{\cal G} \equiv |\Psi_{{00\ldots 0}}\rangle_{\cal G}$.


\subsection{Recurrence scheme}\label{recurrence}

In this section we review the purification protocol introduced in Ref. \cite{Du03} and analyze its properties. We consider two sub--protocols, $P1$ and $P2$, each of which acts on two identical copies $\rho_1=\rho_2=\rho$, $\rho_{12}\equiv\rho_1\otimes \rho_2$. 

\subsubsection{Protocol P1}

In a first step, all parties which belong to the set $V_A$ apply local CNOT operations \cite{noteCNOT} to their particles, with the particle belonging to $\rho_2$ as source, $\rho_1$ as target. Similarly, all parties belonging to set $V_B$ apply local CNOT operations to their particles, but with the particle belonging to $\rho_1$ as source, $\rho_2$ as target. Making use of the properties of graph states, pointed out in Sec. \ref{properties}, together with
\be
{\rm CNOT}=1/2(\eins\otimes\eins+\sigma_z\otimes\eins + \eins\otimes\sigma_x-\sigma_z\otimes\sigma_x), 
\ee
one readily checks that the action of such a multilateral CNOT operations is given by
\be
|\Psi_{{\bm \mu}_{\bf A},{\bm \mu}_{\bf B}}\rangle|\Psi_{{\bm \nu}_{\bf A},{\bm \nu}_{\bf B}}\rangle\rightarrow|\Psi_{{\bm \mu}_{\bf A},{\bm \mu}_{\bf B}\oplus{\bm \nu}_{\bf B}}\rangle|\Psi_{{\bm \nu}_{\bf A}\oplus {\bm \mu}_{\bf A},{\bm \nu}_{\bf B}}\rangle
\label{psitopsi}
\ee
where ${\bm \mu}_{\bf A}\oplus{\bm \nu}_{\bf A}$ denotes bitwise addition modulo 2. For instance, if ${\bm \mu}=\mu_1\mu_2\mu_5$, ${\bm \mu}_{\bf A}\oplus{\bm \nu}_{\bf A}=\mu_1\oplus\nu_1,\mu_2\oplus\nu_2,\mu_5\oplus\nu_5$. 

The second step of protocol $P1$ consists of a measurement of all particles of $\rho_2$, thereby destroying one of the two copies of the initial state. The particles belonging to set $V_A$ are measured in the eigenbasis $\{|0\rangle_x,|1\rangle_x\}$ of $\sigma_x$, while particles belonging to set $V_B$ are measured in the eigenbasis $\{|0\rangle_z,|1\rangle_z\}$ of $\sigma_z$. The measurements in sets $V_A$ [$V_B$] yield results $(-1)^{\xi_j}$ [$(-1)^{\zeta_k}$] respectively, with $\xi_j,\zeta_k \in\{0,1\}$. If the measurement outcomes fulfill $(\xi_j+\sum_{\{k,j\}\in E}\zeta_k){\rm mod}2=0 ~\forall j$ ---which implies ${\bm \mu}_{\bf A}\oplus{\bm \nu}_{\bf A}={\bf 0}$--- the first state is kept. Otherwise, also the first state is discarded and protocol $P1$ failed. In case the resulting state $\tilde \rho$ is kept, one finds that it is again diagonal in the graph--state basis, with new coefficients
\be
\tilde\lambda_{{{\bm \gamma}_{\bf A}},{{\bm \gamma}_{\bf B}}} =\sum_{\{({{\bm \nu}}_{\bf B}, {{\bm \mu}}_{\bf B}) | {{\bm \nu}}_{\bf B} \oplus{{\bm \mu}}_{\bf B}={{\bm \gamma}_{\bf B}}\}} \frac{1}{2K}\lambda_{{{\bm \gamma}_{\bf A}},{{\bm \nu}_{\bf B}}}\lambda_{{{\bm \gamma}_{\bf A}},{{\bm \mu}_{\bf B}}},\label{mapP1}
\ee
where $K$ is a normalization constant such that $\tr(\tilde \rho)=1$ indicating the probability of success of the protocol. 
We note that one may also keep measurement outcomes other than $(\xi_j+\sum_{\{k,j\}\in E}\zeta_k){\rm mod}2=0 ~\forall j$ which would increase the success probability of the protocol. In this case, however, it is not clear whether the modified protocol is still capable of purifying the desired state.

\subsubsection{Protocol P2}

Protocol $P2$ is defined in a similar way and can be obtained from protocol $P1$ by exchanging the roles of sets $V_A$ and $V_B$. The action of the multilateral CNOT operation is in this case given by
\be
|\Psi_{{\bm \mu}_{\bf A},{\bm \mu}_{\bf B}}\rangle|\Psi_{{\bm \nu}_{\bf A},{\bm \nu}_{\bf B}}\rangle\rightarrow|\Psi_{{\bm \mu}_{\bf A}\oplus{\bm \nu}_{\bf A},{\bm \mu}_{\bf B}}\rangle|\Psi_{{\bm \nu}_{\bf A},{\bm \nu}_{\bf B}\oplus{\bm \mu}_{\bf B}}\rangle.
\label{psitopsi2}
\ee 
which leads to new coefficients
\be \tilde\lambda'_{{{\bm \gamma}_{\bf A}},{{\bm \gamma}_{\bf B}}} =\sum_{\{({{\bm \nu}}_{\bf A}, {{\bm \mu}}_{\bf A})| {{\bm \nu}}_{\bf A} \oplus{{\bm \mu}}_{\bf A}={{\bm \gamma}_{\bf A}}\}} \frac{1}{2K}\lambda_{{{\bm \nu}_{\bf A}},{{\bm \gamma}_{\bf B}}}\lambda_{{{\bm \mu}_{\bf A}},{{\bm \gamma}_{\bf B}}},\label{mapP2}
\ee
for the case in which the protocol $P2$ was successful.

\subsubsection{Total purification protocol}

The total entanglement purification protocol is composed of $P1$ and $P2$. It consist in an iterative application of sub--protocols $P1$ and $P2$, always using two identical copies, obtained in the previous round, as input states. It turns out that for certain input states the convergence of the protocol as well as the purification regime can be improved by using an adaptive scheme. That is, instead of using a strictly alternating application of protocols $P1$ and $P2$, one allows for two (or more) subsequent application of the same protocol and may use arbitrary sequences such as $P1-P1-P1-P2-P1-P2-P2-$ etc.. 

We remark that the basic idea of the protocol is similar to the standard recurrence protocols \cite{Be96,De96} for the purification of Bell states. Information about the first state $\rho_1$ is transferred to the second state $\rho_2$ by means of the multilateral CNOT operations and revealed by the measurement. The gain in information about the first state eventually corresponds to an increase of the entanglement of this state. This information transfer becomes evident from Eq.~(\ref{psitopsi}), where we remark that the relevant information is encoded in  ${\bm \mu}_{\bf A},{\bm \mu}_{\bf B}$. One sees that while protocol $P1$ is capable to reveal information about ${\bm \mu}_{\bf A}$, the protocol $P2$ reveals information about ${\bm \mu}_{\bf B}$. In case of a successful purification, the typical action of the total protocol is as follows: The protocol $P1$ increase the weight of  all coefficients $\lambda_{{\bf 0},{\bm \mu}_{\bf B}}$, while $P2$ amplifies coefficients $\lambda_{{\bm \mu}_{\bf A},{\bf 0}}$. In total, this leads to the amplification of $\lambda_{{\bf 0},{\bf 0}}$.

\subsubsection{Binary--like mixtures}\label{binary}

To gain some analytical insight into this procedure, we consider the example of mixed states of the form 
\be
\rho_{\cal A}\equiv \sum_{{\bm \mu}_{\bf A}} \lambda_{{\bm \mu}_{\bf A},{\bf 0}} |\Psi_{{\bm \mu}_{\bf A},{\bf 0}}\rangle \langle \Psi_{{\bm \mu}_{\bf A},{\bf 0}}|.
\label{rhoA}
\ee 
These states arise in a (hypothetical) scenario were all particles within set $V_A$ are only subjected to phase flip errors (described by $\sigma_z$), while all particles within set $V_B$ are subjected to bit flip errors ($\sigma_x$), which can also be described as a collection of phase flip errors in set $V_A$ (see Sec. \ref{properties}). We remark that this situation is equivalent to a more natural scenario where only phase flip errors occur on all locations and one considers a state which is up to local unitary operations equivalent to $\rho_{\cal A}$. Such a situation may for instance occur when each of the particles of a perfect TCGS is subjected to decoherence described by a dephasing quantum channel.

From the discussion in the previous section, it is clear that the iterative application of protocol $P1$ is sufficient to purify states of the form Eq. (\ref{rhoA}), as only information about ${\bm \mu}_{\bf A}$ has to be extracted. A single application of protocol $P1$ leads again to a state of the form $\rho_{\cal A}$, with new coefficients 
\be
\tilde\lambda_{{\bm \mu}_{\bf A},{\bf 0}} = \lambda_{{\bm \mu}_{\bf A},{\bf 0}}^2/K,\label{binary1}
\ee
where $K=\sum_{{\bm \mu}_{\bf A}} \lambda_{{\bm \mu}_{\bf A},{\bf 0}}^2$ is a normalization constant which gives the probability of success of the protocol. That is, the largest coefficient is amplified with respect to the other ones. It follows that iteration of the protocol allows one to produce pure graph states $|\Psi_{{\bf 0},{\bf 0}}\rangle$ with arbitrary high accuracy, given the coefficient $\lambda_{{\bf 0},{\bf 0}}$ is larger than all other coefficients $\lambda_{{\bm \mu}_{\bf A},{\bf 0}}$. That is, the condition that successful purification is possible reads $\lambda_{{\bf 0},{\bf 0}} > \lambda_{{\bm \mu}_{\bf A},{\bf 0}} \forall{\bm \mu}_{\bf A}\not={\bf 0}$. If this conditions is fulfilled, the protocol converges towards the attracting fixed point given by $\lambda_{{\bf 0},{\bf 0}}=1$. If not, we choose the largest coefficient, say $\lambda_{{\bm \mu}_{\bf A},{\bf 0}}$, and map it onto $\lambda_{\bf 0,0}$ via local unitary operations. We remark that the family of states $\rho_{\cal A}$ includes states up to rank $2^{N_A}$, which ---depending on the corresponding graph--- can be as high as $2^{N-1}$.

As a concrete example, consider the one parameter family $\rho_{\cal A}(F)$ with $\lambda_{{\bf 0},{\bf 0}}=F$, $\lambda_{{\bm \mu}_{\bf A},{\bf 0}}=(1-F)/(2^{N_A}-1)$ for ${\bm \mu}_{\bf A} \not={\bf 0}$, where $F$ is the fidelity of the desired state. Application of protocol $P1$ keeps the structure of those states and leads to 
\be
\tilde F = \frac{F^2}{F^2+ (1-F)^2/(2^{N_A}-1)}.
\ee
This map has $\tilde F=1$ as attracting fixed point for $F\geq 1/2^{N_A}$. The probability of success for a single step is given by 
$p=F^2+ (1-F)^2/(2^{N_A}-1)$.


\subsubsection{Purification regime and convergence}

While for the restricted family of states $\rho_{\cal A}$ discussed in the previous section an analytic treatment of the protocol is possible, the situation is more complicated in the general case. For full rank mixed states, an iterative application of both protocols, $P1$ and $P2$, is required to reveal information about ${\bm \mu}_{\bf A}$ and ${\bm \mu}_{\bf B}$ respectively. In this case, the action of each protocol is described by a more complicated non--linear mapping (see Eqs. (\ref{mapP1},\ref{mapP2})) of a large number of independent variables (in total $2^N-1$) which makes an analytic treatment of the protocol very difficult. We have not been able to determine boundaries of the purification regime and the convergence properties of the protocol analytically in the general case. For a large family of states, arising from different noise models, we have however investigated the purification regime and convergence properties numerically.

As a first example, we consider noisy TCGS arising naturally in a multiparty communication scenario where each of the $N$ particles constituting $|\Psi_{{\bf 0}}\rangle$ is sent through a noisy quantum channel. We consider depolarizing channels with noise parameter $q$ described by
\be
{\cal E}_k\rho= q \rho +(1-q)/2\eins_k \otimes \tr_k(\rho), \label{whitenoise}
\ee
where the channel is acting on particle $k$. The resulting multipartite state is of the form 
\be
\rho(q) = {\cal E}_1{\cal E}_2 \ldots {\cal E}_N |\Psi_{{\bf 0}}\rangle\langle \Psi_{{\bf 0}}|.
\ee
We point out that $q=1$ corresponds to perfect transmission, i.e. no decoherence, while $q=0$ leads to a completely depolarized state. We have numerically investigated the threshold value $q_{\rm min}$ until which our multiparticle entanglement purification protocol can be successfully applied. For $q\geq q_{\rm min}$, we have that the purification protocol can be successfully applied, while for $q<q_{\rm min}$ the protocol fails. The results of this numerical investigation are summarized in Fig. \ref{Fig_Fminqmin} for linear cluster states and GHZ states of different size. While for linear cluster states one observes that the threshold value $q_{\rm min}$ is essentially independent of the number of particles $N$, the situation for GHZ states is different. For GHZ states the threshold value $q_{\rm min}$ increases with the number of qubits. That is, the tolerable amount of white noise per particles decreases with increasing $N$ and thus it becomes more difficult to purify large scale GHZ states. We have also analyzed other two-colorable graph states and found that the threshold value does in general not depend on the size of the system $N$, but is determined by the maximal degree of the corresponding graph. For specific families of states, the degree of the graph may however depend on the number of vertices. An example is given by the $N$--particle GHZ state, where the degree of the corresponding graph is $N-1$, i.e. the degree scales with the size of the system. 
Indeed, it can be shown analytically \cite{Du03a} that the value of $q$ such that GHZ states become non--distillabe (by any protocol) increases with increasing $N$. For families of graph states of fixed degree and arbitrary size, however, one can show that the states remain distillable if $q\geq q_{\rm crit}$, where $q_{\rm crit}$ only depends on the degree of the graph. This different behavior can be intuitively understood as follows: Consider only bit flip errors described by $\sigma_x$. If the degree of the graph is high, a certain vertex is connected to a large number of neighboring vertices. Whenever a bit flip error in one of the neighboring vertices occurs, this is equivalent to a phase flip error (described by $\sigma_z$) at the vertex in question as can be seen from the discussion in Sec. \ref{properties}. That is, a large number of independent errors affect a single vertex (and thus a specific index $\mu_j$) and these errors accumulate, leading to a threshold value increasing with the degree of the graph \cite{footnotevertex}.   
We remark that whenever $q \geq q_{\rm min}$, our protocol successfully converges towards the fixed point specified by $\lambda_{{\bf 0},{\bf 0}}=1$.

Note that the different behavior of GHZ states and graph states with fixed degree is not reflected by the minimal required fidelity $F_{\rm min}\equiv\langle\Psi_{\bf 0,{\bf 0}}|\rho(q_{\rm min})|\Psi_{{\bf 0},{\bf 0}}\rangle$ which is in both cases decreasing exponentially with the size of the system $N$. For linear cluster states and GHZ states, $F_{\rm min}$ is plotted in Fig.~\ref{Fig_Fminqmin} for different number of particles $N$. These observations suggest that the fidelity is for multiparticle system not a very sensitive measure to judge properties of multiparticle entangled states in the presence of decoherence. From the exponential decrease of the minimal required fidelity, one would be tempted to conclude that the requirement to purify states become less stringent with increasing size of the system. This is however certainly not true, as the tolerable amount of white noise per particle may even decrease with the size of the system, e.g. for GHZ states.

\begin{figure}[ht]
\begin{picture}(230,200)
\put(-5,-5){\epsfxsize=230pt\epsffile[15 169 550 588]{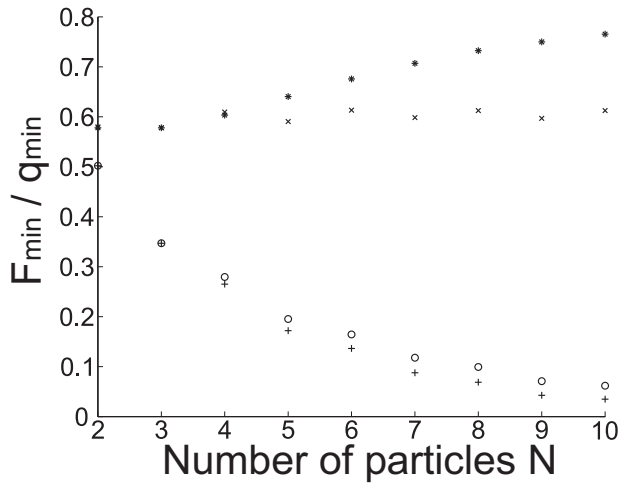}}
\end{picture}
\caption[]{(a): Minimal value of fidelity $F_{\rm min}$ $+$ [$\circ$] and parameter $q_{\rm min}$ $\times$ [$*$] for linear cluster states [GHZ states] for different number of particles $N$ and perfect local operations.}
\label{Fig_Fminqmin}
\end{figure}

We have also considered mixed states of the form 
\be
\rho(x)=x|\Psi_{{\bf 0}}\rangle\langle \Psi_{{\bf 0}}|+(1-x) /2^N\eins,\label{GWerner}
\ee
i.e.\ mixtures of the desired state with a completely depolarized state. We observe that the situation is similar as in the case of local white noise, i.e.\ $F_{\rm min}\equiv x_{\rm min}+(1-x_{\rm min})/2^N$ decreases exponentially with $N$. For $x\geq x_{\rm min}$, the protocol successfully converges and produces perfect two--colorable graph states. The threshold value $F_{\rm min}$ is plotted for linear cluster states of different size in Fig. \ref{fig:min_fidelity}. For $n=2$ and $n=3$, the minimum required fidelity coincides with the values found for the purification of GHZ-states \cite{Mu99}. The reason for this coincidence is that
the two- and tree-party linear cluster states are (up to local unitary operations) equal to two- and tree-qubit GHZ states, and that the cluster purification protocol is (in these two cases) equivalent to
the GHZ purification protocol.

\begin{figure}[ht]
\begin{picture}(230,190)
\put(-5,-5){\epsfxsize=230pt\epsffile[19 444 337 711]{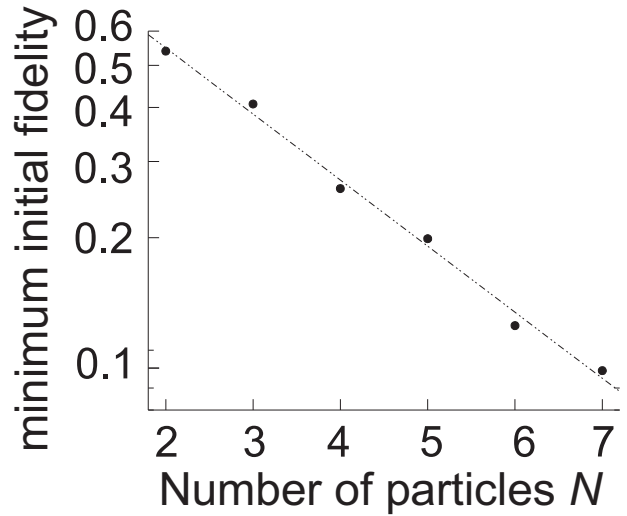}}
\end{picture}
\caption[Minimum fidelity in the purification protocol]
  {The required initial fidelity for linear cluster states as a function of the number $N$ of parties.
  The dotted curve is an exponential fit to the
    exact values (circles).} 
  \label{fig:min_fidelity}
\end{figure}

For more general states, the purification regimes as well as the convergence of the protocol is difficult to determine due to the large number of parameters.

\subsubsection{Efficiency and yield}

The recurrence scheme presented in the previous section is capable of purifying a large class of possible input states. As in the bipartite case, however, the protocol approaches unit fidelity (and thus successful perfect distillation) only in the asymptotic limit, i.e. a large number of iterations of the protocol is required. Since any step of the protocol only succeeds with certain probability, and in addition one pair is consumed in each step regardless of the measurement outcomes, the recurrence protocol has ---strictly speaking--- zero yield. Here, the yield of the protocol is defined as the number of 
copies of the state which are, on average, required to produce a single copy of the desired (pure) output state.
For practical purposes it is often sufficient to produce output states with a fidelity larger than a certain threshold value and thus a finite, possibly small number of iteration steps suffices. The efficiency of the procedure achieving this task can be easily evaluated. For a single iteration of the entanglement purification protocol one obtains that the average number of copies required to obtain a single copy of the output state is given by $2/K$, where $K$ is the probability of success of the protocol (see e.g. Eq. (\ref{mapP1}) for protocol $P1$). The efficiency of this purification step is thus given by $K/2$. Note that if the fidelity of the initial state approaches unity, we have that $K \to 1$. The yield of the total procedure is obtained by multiplying the efficiencies of the individual purification steps. In Sec. \ref{comparebipartite}, the efficiency of multiparticle entanglement purification protocol will be compared to the efficiency of protocols based on bipartite entanglement purification.

We remark that the yield of the purification protocol decreases (exponentially) with the number of parties $N$, as the probability of success for each purification step, $K$ (see e.g. Eq. (\ref{mapP1})), decreases with $N$. To overcome practical difficulties for states consisting of large number of particles (where the success probability may be very small), it is possible to use an alternative purification method which essentially consists in a cut and re--connect procedure. That is, a given TCGS is split up by means of local measurements into several (smaller) sub--graph states. These sub--graph states are then purified independently and finally these sub--graph states are re--connected. Since the sub--graphs are smaller, the yield for the purification of each sub--graph state is higher. The re--connection is deterministic and may e.g. be performed by means of Bell--type measurements, where sub--graphs are chosen in such a way that each of them is itself a two--colorable graph (and hence distillable by our protocol) and the sub--graphs overlap at the re--connection points (so that the Bell--type measurements are in fact local operations). As several (non--overlapping) sub--graph states can be produced from a single copy of the initial graph state, the yield of the total procedure is essentially determined by the yield to purify the sub--graph states. Extremal cases of this procedure are on the one hand the purification of pairs and creation of the required target state by means of teleportation and on the other hand direct multiparty purification, each of which having its own advantages and disadvantages. The optimal choice of the size of the subgraph will depend on the required task. Optimization can be performed with respect to the yield, the achievable fidelity and the purification regime and will be treated elsewhere.

\subsection{Hashing and breeding}\label{hashing}

It is interesting from a principal point of view to obtain purification protocols which have non--zero yield. In the bipartite case, the hashing and breeding protocol (see Ref. \cite{Be96}) achieve this aim. In these protocols, the local operators act jointly on a large number $M$ of copies of an initial state $\rho$, where $M\rightarrow \infty$. In brief, they use entanglement ---either present in pure form (breeding) or in noisy form (hashing)--- to reveal (non--local) information about $\rho^{\otimes M}$. This information gain results in purification of a certain subensamble of $M'$ copies. The yield in the case of hashing is given by $M'/M$, while in the case of breeding one has to take into account that entangled pure states consumed during the purification procedure have to be given back.

The hashing protocol has been generalized by Maneva and Smolin \cite{Sm00} to a multipartite setting. They showed that certain multiparty entangled states, namely GHZ states, can be purified. To be specific, the protocol introduced in Ref. \cite{Sm00} allows one to purify states diagonal in the basis of GHZ states with a non--zero yield, provided the initial fidelity of the state is sufficiently high. In this section, we will show that the hashing protocol of Maneva and Smolin can be generalized to purify a much larger class of possible output states. In particular, we will present for each two colorable graph state a protocol which is capable to produce this graph state as an output state with non--zero yield, provided the initial fidelity is sufficiently high. The main point is to realize that the stabilizer formalism used in Ref. \cite{Sm00} to construct a purification protocol for GHZ states can be applied in a similar way to two-colorable graph states. In fact, Eqs. (\ref{psitopsi},\ref{psitopsi2})  which describe the action of certain multilateral CNOT operations on two graph states already show how information about an unknown graph state can be transferred from one copy to another. This information can be revealed by measurements. In particular, the whole bit string ${\bm \mu}_{\bf A}$ of a single copy of a two--colorable graph state $|\Psi_{{\bm \mu}_{\bf A},{\bm \mu}_{\bf B}}\rangle$ can be obtained by performing a local measurement in the eigenbasis of $\sigma_x$ of all particles in set $V_A$, while all particles in set $V_B$ are measured in the eigenbasis of $\sigma_z$. The measurements in sets $V_A$ [$V_B$] yield results $(-1)^{\xi_j}$ [$(-1)^{\zeta_k}$] respectively, with $\xi_j,\zeta_k \in\{0,1\}$. The value of the bit $\mu_j$, $j\in V_A$ is given by
\be
\mu_j=\left(\xi_j+\sum_{\{k,j\}\in E}\zeta_k\right){\rm mod}2,
\ee 
which follows from the eigenvalue equation Eq. (\ref{EV}). That is, the measurements allow one to simultaneously determine the eigenvalues of all correlation operators $K_j$ for $j\in V_A$. In a similar way, by exchanging the role of $V_A$ and $V_B$, one can obtain the bit string ${\bm \mu}_{\bf B}$. Note, however, that bit strings ${\bm \mu}_{\bf A}$ and ${\bm \mu}_{\bf B}$ can not be determined simultaneously by {\em local} measurements. 

Given these tools, the hashing (and Breeding) protocol can now be implemented in the usual manner. That is, given $M$ copies of a mixed state $\rho$ diagonal in the graph state basis (which can always be achieved by applying the depolarization procedure described in Sec. \ref{depolarization})
\be
\rho=\sum_{{\bm \mu}_{\bf A},{\bm \mu}_{\bf B}} \lambda_{{\bm \mu}_{\bf A},{\bm \mu}_{\bf B}} |\Psi_{{\bm \mu}_{\bf A},{\bm \mu}_{\bf B}}\rangle\langle \Psi_{{\bm \mu}_{\bf A},{\bm \mu}_{\bf B}}|,
\ee
one chooses a random subset of $m$ copies and determines the parity of each bit $\mu_j$. This can be accomplished by applying multilateral CNOT operations between the first $m-1$ copies of the set and the $m^{\rm th}$ copy. The corresponding measurement of the $m^{\rm th}$ copy allows to determine the parity of the whole bit string ${\bm \mu}_{\bf A}$ of the $m-1$ remaining copies. The procedure is repeated for many of these randomly chosen subsets and in a similar way the parity of ${\bm \mu}_{\bf B}$ is determined for other random subsets. It is now straightforward to calculate the number of required repetitions of the above procedure to determine completely all relevant information of the remaining copies. To this aim, we define the coefficients $a_j^{(0)}, a_j^{(1)}$ as follows:
\be
a_j^{(\mu_j)}= \sum_{\mu_k \not= \mu_j} \lambda_{\mu_1\mu_2 \ldots \mu_j \ldots \mu_N}.
\ee
For instance, for $N=3$ we have that $a_1^{(0)}=\sum_{k,l}\lambda_{0kl}, a_1^{(1)}=\sum_{k,l}\lambda_{1kl}$ while $a_3^{(0)}=\sum_{i,j}\lambda_{ij0}$ and $a_j^{(0)}+a_j^{(1)}=1$. The entropy $S(a_j^{(0)},a_j^{(1)})$ is given by
\be
S(a_j^{(0)},a_j^{(1)})=-a_j^{(0)}\log_2a_j^{(0)}-a_j^{(1)}\log_2a_j^{(1)}
\ee
and determines the number of copies which has to be measured in order to obtain bit $\mu_j$. Following the reasoning of Refs. \cite{Be96, Sm00}, we can now determine the yield of the hashing protocol and find 
\bea
D=1&-&{\rm max}_{j \in V_A} [\{S(a_j^{(0)},a_j^{(1)})\}]\nonumber\\
 &-& {\rm max}_{k \in V_B} [\{S(a_k^{(0)},a_k^{(1)})\}]
\eea
For mixed states of the form Eq. (\ref{GWerner}), that are mixtures of a pure graph state with the maximally mixed state, we have that $a_j^{(0)}=(1+x)/2, a_j^{(1)}=(1-x)/2 \forall j$. The yield of the protocol is in this case given by 
\be
D=1-2 S\left(\frac{1+x}{2}, \frac{1-x}{2}\right).
\ee
Note that the yield of the hashing protocol approaches one for any state diagonal in the graph state basis which fulfills $\lambda_{{\bf 0}} \rightarrow 1$, independent of the specific form of the state. In particular, this implies that if a given mixed state has sufficiently high fidelity $F$, the hashing protocol (combined with the depolarization procedure) allows one to extract pure two--colorable graph states with non--zero yield, and the yield approaches one for $F\rightarrow 1$.


\section{Imperfect local operations}\label{ILO}


Until now, we have assumed that local operations ---in particular CNOT operations--- are perfect. In practice, however, these operations as well as measurements will be imperfect. We now investigate the influence of errors in the local operations on the multiparticle entanglement purification protocol.  We will consider an error model where imperfect local two--qubit operations are described by the completely positive map
\be
{\cal E}_{U_{jk}}\rho = U_{jk} [{\cal E}_j {\cal E}_k\rho] U_{jk}^{\dagger},\label{noisyU}
\ee
where ${\cal E}_k, {\cal E}_j$ are given by Eq. (\ref{whitenoise}) with error parameter $p$. That is, an imperfect operation is described by first applying local white noise with probability $(1-p)$ independently on the qubits, followed by the perfect unitary operation. Such an error model allows us to analyze the protocol up to $N=13$, involving $2N=26$ qubits. For smaller number of particles, we have also investigated more general error models, e.g. two--qubit correlated white noise, and also errors in the measurement process, observing essentially the same behavior as for this simple model. 

We have numerically investigated the dependence of the minimal required fidelity and the maximal reachable fidelity for linear cluster states of different length on error parameters $p$ (see Fig. \ref{Fig_FminFmax}). We remark that whenever the fidelity of the initial state (which is obtained from a perfect cluster state by applying local white noise with a certain noise parameter) fulfills $F_{\rm min} \leq F \leq F_{\rm max}$, the entanglement purification protocol converges towards a state with $F=F_{\rm max}$. That is, for any given error parameter $p$, $F_{\rm min}$ and $F_{\rm max}$ determine the purification regime where our protocol can be successfully applied in order to increase the fidelity of the state. As can be seen from Fig. \ref{Fig_FminFmax}, the purification regime becomes broader with increasing $N$. In particular, the minimal value of $p$ such that a finite purification regime remains, i.e. the threshold value $p_{\rm min}$ until which our MEPP can be successfully applied, is (almost) independent of the number of parties $N$ which can be seen from Fig. \ref{Fig_thresholdp}. It even seems that for larger number of particles the tolerable amount of noise per operation is larger. 
Performing a similar investigation for GHZ states, we find on the contrary that the threshold value $p_{\rm min}$ {\em increases} with increasing $N$ (Fig. \ref{Fig_thresholdp}), i.e. it becomes more difficult to purify GHZ states with a large number of particles $N$.

\begin{figure}[ht]
\begin{picture}(230,200)
\put(-5,-5){\epsfxsize=230pt\epsffile[ 1 141 536 578]{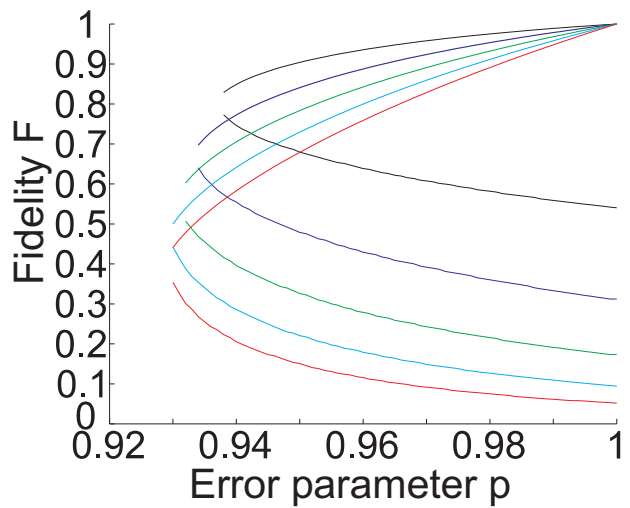}}
\end{picture}
\caption{Maximal reachable fidelity $F_{\rm max}$ and minimal required fidelity $F_{\rm min}$ plotted against error parameter $p$ (local operations) for density operators arising from single-qubit white noise. Curves from top to bottom correspond to linear cluster states with $N=2,4,6,8,10$ particles.} 
\label{Fig_FminFmax}
\end{figure}  

\begin{figure}[ht]
\begin{picture}(230,200)
\put(-5,-5){\epsfxsize=230pt\epsffile[6 155 577 592]{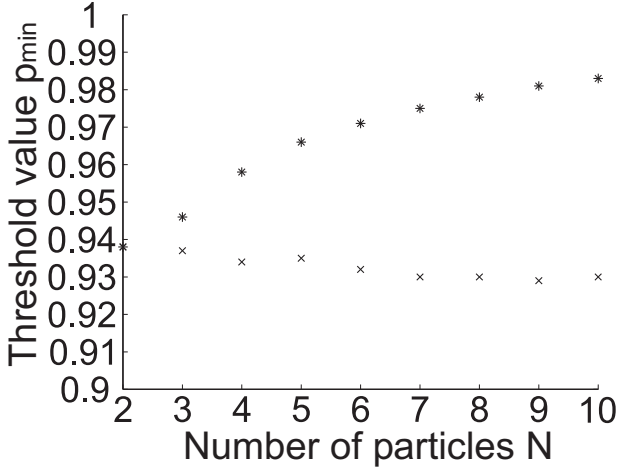}}
\end{picture}
\caption{Threshold value for errors in local operations $p_{\rm min}$ for GHZ states ($*$) and linear cluster states ($\times$) with different number of particles $N$.} 
\label{Fig_thresholdp}
\end{figure}  







\section{Purification regime for binary--like mixtures}\label{binarymix}

As in the case of perfect local control operations, it is possible to treat binary--like mixtures $\rho_{\cal A}$ of the form Eq. (\ref{rhoA}) analytically when considering a restricted error model which keeps the structure of these states. Note that considering such an error model with this restricted kind of errors allows one to obtain a lower bound on the threshold value for more general error models. To this aim, we consider the completely positive map (CPM) 
${\cal M}_j$ given by
\be
{\cal M}_j\rho= p \rho +\frac{1-p}{2} (\rho+ \sigma_x^{(j)} \rho \sigma_x^{(j)}),\label{calM}
\ee
which correspond to a bit--flip channel acting on qubit $j$. We model imperfect local unitary operations by the following CPM
\be
{\cal E}_{U_{jk}}\rho = U_{jk} [{\cal M}_j {\cal M}_k\rho] U_{jk}^{\dagger},\label{calE}
\ee
where ${\cal M}_j$ is given by Eq. (\ref{calM}) if qubit $j$ belongs to the set $V_B$, and the identity otherwise. That is, we assume that operations on particles in set $V_A$ are perfect, while an imperfect unitary operation acting on two qubits held by a party in set $V_B$ is described by first applying a probabilistic bit--flip channel on the qubits, followed by the ideal unitary operation. Such an error model ensures that the structure of binary--like mixtures (Eq. (\ref{rhoA})) is maintained. In principle, one could in addition also consider phase flip errors for all particles in set $V_A$ ---which would still maintain the structure of binary mixtures---, however the analysis is more complex and no additional insight is gained. 

In the following, we will investigate the purification regime for GHZ states and closed linear cluster states, initially of the form $\rho_{\cal A}(F)$. That is, we will determine the threshold value $p_{\rm crit}$ until which a single instance of our purification protocol allows one to increase the quality of the state. While we find that for closed linear cluster states the threshold value $p_{\rm crit}$ essentially remains constant, independent of the size of the system, for GHZ states we show that even for this restricted kind of errors, the threshold value increases with $N$, approaching 1 in the limit of large $N$. This implies that purification of GHZ states with large number of particles becomes exceedingly difficult with increasing $N$. In the limit of large $N$, nearly noiseless local operations are required. On the contrary, the requirements on local operations for the purification of cluster states is independent of the number of particles $N$.  


\subsection{GHZ states}

We start by investigating the properties of binary--like mixtures of GHZ states. Recall that the corresponding graph of a GHZ state is given by edges $\{1,k\}$, $k\in\{2,3,\ldots,N\}$ and $V_A=\{V_1\}$, $V_B=\{V_2,V_3,\ldots, V_N\}$. We consider states of the form
\be
\rho_{\cal A}(x)=x |\Psi_{0,{\bf 0}}\rangle\langle\Psi_{0,{\bf 0}}| + (1-x)/2 \eins_{V_A},\label{binGHZ}
\ee 
where $\eins_{V_A}=|\Psi_{0,{\bf 0}}\rangle\langle\Psi_{0,{\bf 0}}|+|\Psi_{1,{\bf 0}}\rangle\langle\Psi_{1,{\bf 0}}|$. As pointed out in Sec. (\ref{properties}), the action of a bit flip error $\sigma_x$ on any of the particles $2,3,\ldots, N$ on graph states can equivalently be described by a phase flip error $\sigma_z$ on particle 1. In particular we have that for $j=2,3,\ldots ,N$, ${\cal M}_j^{(B)} |\Psi_{0,{\bf 0}}\rangle\langle\Psi_{0,{\bf 0}}|  = p |\Psi_{0,{\bf 0}}\rangle\langle\Psi_{0,{\bf 0}}|  + (1-p)/2 \eins_{V_A}$ and also ${\cal M}_j^{(B)}\eins_{V_A}=\eins_{V_A}$. It readily follows that the action of the purification protocol $P1$ which involves imperfect unitary operations on two copies of the input state $\rho_{\cal A}(x)$ can equivalently be described by the action of the perfect protocol $P1$ on two copies of the state $\tilde \rho_{\cal A}(x') \equiv {\cal M}_2^{(B)}{\cal M}_3^{(B)}\ldots {\cal M}_N^{(B)}\rho_{\cal A}(x)$. One finds that 
\be
\tilde \rho_{\cal A}(x')= \rho (x p^{N-1}),
\ee 
that is the state is still of the form Eq. (\ref{binGHZ}) with new coefficient $x'=xp^{N-1}$. The action of the perfect protocol $P1$ on $\tilde \rho_{\cal A}(x')$ is given by Eq. (\ref{binary1}) with $\lambda_{0,{\bf 0}}=x'+(1-x')/2, \lambda_{1,{\bf 0}}=(1-x')/2$ yielding
\be
\tilde\lambda_{0,{\bf 0}}=[xp^{N-1}+\frac{1-xp^{N-1}}{2})]^2/K.
\ee
The purification protocol was successful if the fidelity of the resulting state $\tilde\lambda_{0,{\bf 0}}$ is larger than the one of the initial state $\rho(x)$, $F\equiv \lambda_{0,{\bf 0}}=x+(1-x)/2$. That is
\be
\frac{[xp^{N-1}+\frac{1-xp^{N-1}}{2}]^2}{[xp^{N-1}+\frac{1-xp^{N-1}}{2}]^2+[\frac{1-xp^{N-1}}{2}]^2} \geq x+\frac{1-x}{2},
\ee
which can be rewritten as
\be
2p^{N-1}-1 \geq x^2 p^{2(N-1)}.\label{condGHZ}
\ee 
On the one hand, for a fixed noise level of local operations (given by the error parameter $p$)  Eq. (\ref{condGHZ}) allows one to obtain the maximal reachable fidelity $F_{\rm max} \equiv x_{\rm max}+(1-x_{\rm max})/2$, that is the fixed point of the protocol. One finds 
\be
x_{\rm max} = \sqrt{(2p^{N-1}-1})/p^{(N-1)}.
\ee
On the other hand, one can also determine the threshold value for the error parameter $p$, $p_{\rm crit}$, i.e. the minimum required reliability of the local operations that purification is possible. For $(2p^{N-1}-1) < 0$, the inequality (\ref{condGHZ}) can certainly not be fulfilled, independent of $x$. Thus independent of the initial quality of the state, the protocol is not capable to increase the fidelity if $p<p_{\rm crit}$. A lower bound on the threshold value $p_{\rm crit}$ is thus given by
\be
p_{\rm crit}= \left(\frac{1}{2}\right)^{1/(N-1)},
\ee 
which increases for increasing $N$. That is, even if we consider only a restricted kind of errors on particles within set $V_B$, the requirements on the quality of local operations become more stringent if the number of particles $N$ increases. This is in agreement with the numerical results found for the more general white noise error model discussed in the previous section.


\subsection{Closed linear cluster states}

We now turn our attention to closed linear cluster states of size $N\equiv 2 M$, specified by a graph with $N$ vertices and edges $\{k,(k+1){\rm mod} N\}$. The sets $V_A$ [$V_B$] are given by all odd [even] vertices respectively. As in the case of GHZ states we determine not only the minimal required and maximal reachable fidelity, but also the threshold values for local operations. We find that the tolerable amount of noise per imperfect two--qubit operation essentially remains constant independent of the number of particles involved, and is for large $N$ given by $p_{\rm crit}\approx 0.4976$. That is, the purification protocol is also for large number of particles remarkable robust against the influence of imperfect local operations, which is interesting for possible practical applications.

We consider density operators of the form 
\be
\rho_{\cal A}(x)\equiv x|\Psi_{{\bf 0},{\bf 0}}\rangle\langle\Psi_{{\bf 0},{\bf 0}}|+\frac{1-x}{2^{N_A}}  \eins_{V_A},
\ee
where $\eins_{V_A}\equiv \sum_{{\bm \mu}_{\bf A}}  |\Psi_{{\bm \mu}_{\bf A},{\bf 0}}\rangle \langle \Psi_{{\bm \mu}_{\bf A},{\bf 0}}|$.
We have that $\rho_{\cal A}(x)$ has rank $2^{N_A}=2^{M}$ and the fidelity $F$ of the state with respect to $|\Psi_{{\bf 0},{\bf 0}}\rangle$ is given by $F=x+(1-x)/2^{M}$. For simplicity, we will assume $M$ odd in our analysis. A similar analysis can be performed for $M$ even. We will consider the purification protocol $P1$, which is sufficient to purify these kind of states. We analyze a single instance of the purification protocol $P1$ and determine the conditions under which an increase of the Fidelity $F$ is possible. Recall that imperfect local unitary operations are modelled by Eq. (\ref{calE}). It turns out to be convenient to use the parameter $q \equiv (1+p)/2$ to describe the quality of imperfect local operations (see Eq. (\ref{calM})).

As in the case of GHZ states, the action of the imperfect protocol $P1$ on two copies of the state $\rho_{\cal A}(x)$ can be equivalently described by the action of the {\em perfect} (error free) protocol $P1$ on two copies of an input state $\rho'_{\cal A}$. We have that 
\be
\rho'_{\cal A} \equiv {\cal M}_1{\cal M}_2\ldots {\cal M}_N\rho_{\cal A}(x),
\ee
where ${\cal M}_k$ is defined in Eq. (\ref{calM}) for $k\in V_B$ and is given by the identity operation if $k\in V_A$. It is relatively straightforward to determine $\rho'_{\cal A}$. Using that ${\cal M}_1{\cal M}_2\ldots {\cal M}_N \eins_{V_A}=\eins_{V_A}$, it only remains to determine the action of ${\cal M}_1{\cal M}_2\ldots {\cal M}_N$ on the cluster state $|\Psi_{{\bf 0}}\rangle\langle \Psi_{{\bf 0}}|$. Since the action of $\sigma_x$ on particle $k$ of a cluster state can be equivalently described by $\sigma_z$ operations on the neighboring particles $k-1$ and $k+1$, we have that the resulting state ${\cal M}_1{\cal M}_2\ldots {\cal M}_N|\Psi_{{\bf 0}}\rangle\langle \Psi_{{\bf 0}}|$ is again diagonal in the graph state basis, where only some of the coefficients $\alpha_{{\bm \mu}_{\bf A},{\bf 0}}$ are non--zero. A straightforward calculation shows that for $M$ odd one obtains a total of $2^{M-1}$ non--zero terms with corresponding coefficients $\{\alpha_k\}$, where $0\leq k \leq (M-1)/2$ and $\alpha_k$ appears $b_{M,k}$ times, where 
\be
b_{M,k}\equiv \left(\begin{array}{c} M \\ k \end{array}\right)= M!/(k!(M-k)!).
\ee
We have that $\alpha_k$ is given by
\be
\alpha_k\equiv q^k(1-q)^{M-k}+q^{M-k}(1-q)^k,
\ee
where $\alpha_0$ corresponds to $|\Psi_{{\bf 0}}\rangle\langle\Psi_{{\bf 0}}|$. 

That is, the state $\rho'_{\cal A}$ is diagonal in the graph state basis with coefficients $\lambda'_{{\bm \mu}_{\bf A},{\bf 0}}$. These coefficients are given by
\bea
\lambda'_{k}&=&x[q^k(1-q)^{M-k}+q^{M-k}(1-q)^k]+\frac{1-x}{2^{M}},\\
\lambda'_{M}&=&\frac{1-x}{2^{M}},
\eea
where $0\leq k \leq (M-1)/2$. Each of the coefficients $\lambda'_k$ appears $b_{M,k}$ times, while the coefficient $\lambda'_M$ appears $2^{M-1}$ times. Note that $\lambda'_0$ corresponds to $\lambda'_{{\bf 0}}$, i.e. determines the fidelity of the state $\rho'_{\cal A}$. 

The action of the (perfect) purification protocol $P1$ is given by Eq. (\ref{binary1}) and can be determined straightforwardly. In particular, the fidelity $F$ of the resulting state after a successful purification step is given by   
\be
\tilde \lambda_0 = (\lambda'_0)^2/\Gamma,
\ee
with 
\be
\Gamma=\sum_{k=0}^{(M-1)/2} b_{M,k} (\lambda'_k)^2 + 2^{M-1} (\lambda'_M)^2.\label{Gamma}
\ee
The imperfect purification protocol is capable to increase the fidelity if $\tilde\lambda_0 > \lambda_0$, where $\lambda_0=x+(1-x)/2^M$. To evaluate the sums appearing in Eq. (\ref{Gamma}) one only need to realize the following identity
\bea
\sum_{k=0}^{(M-1)/2}[b_{M,k}q^k(1-q)^{M-k}+ b_{M,k}q^{M-k}(1-q)^{k}]=\nonumber\\
=\sum_{k=0}^{M}b_{M,k}q^k(1-q)^{M-k}.
\eea
The resulting binomial sums can then be easily evaluated and one finds e.g.  
\bea
\sum_{k=0}^{M}b_{M,k}q^{2k}(1-q)^{2M-2k}&=&(1-q)^{2M}\left[1+\left(\frac{q}{1-q}\right)^2\right]^M\nonumber\\
\sum_{k=0}^{(M-1)/2}b_{M,k}&=&2^{M-1}.
\eea
It turns out to be useful to define the functions $A\equiv A(q),B\equiv B(q),C\equiv C(q)$ given by
\bea
A&=&q^M+(1-q)^M-\frac{1}{2^M}\nonumber\\
B&=&\frac{1}{2^M}\\
C&=&(1-q)^{2M}\left[1+\left(\frac{q}{1-q}\right)^2\right]^M-\frac{1}{2^M}+[2q(1-q)]^M.\nonumber
\eea
After some algebra, one finds that $\Gamma=x^2C+B$ and $\tilde\lambda_0 = [xA+B]^2/\Gamma$. The condition that a single successful application of the imperfect purification protocol $P1$ leads to an increase of the fidelity is thus given by
\be
\frac{[xA+B]^2}{x^2C+B} \geq x(1-B)+B.
\ee
The corresponding purification regime can be determined by solving the resulting quadratic equation in $x$. One obtains 
\be
x_\pm=\frac{BC-A^2 \pm \sqrt{\Delta}}{C(B-1)},\label{xpm}
\ee
with
\be
\Delta=(A^2-BC)^2+4C(1-B)[2AB-B(1-B)].
\ee 
That is, for $x_- \leq x \leq x_+$ a successful purification (resulting in an increase of the fidelity of the state) is possibly. Recall that $x_-,x_+$ are functions of $q$, so Eq. (\ref{xpm}) determines the purification regime for any fixed error parameter $q=(1+p)/2$. For instance, if $q=0.9$ a single application of the protocol $P1$ increase the fidelity $F\equiv x+(1-x)/2^{N/2}$ in the range $2^{-0.33 N}\leq x \leq 2^{-0.009 N}$. That is, for each $N$ there exists a finite regime where entanglement purification is possible. The threshold value $q_{\rm crit}$ (respectively $p_{\rm crit}$) until which successful purification is possible for some input states can be determined by (numerically) solving the polynomial equation $\Delta =0$. One finds that the threshold value $q_{\rm crit}$ [$p_{\rm crit}$] slightly varies over $N$ in the interval $0.7001 \leq q \leq 0.7491$ and converges for large $N$ towards $q_{\rm crit} \approx 0.7469$ [$p_{\rm crit}\approx 0.4938$] (see Fig. (\ref{Fig_thresholdp_anal})). That is, independent of the size of the cluster state, the tolerable amount of noise for local operations specified by $q$ remains (approximately) constant and approaches a finite value $q_{\rm crit}^{\infty} \not=1$. This is in contrast to the behavior of GHZ states but confirms the numerical results found for the more general error model of white noise.

\begin{figure}[ht]
\begin{picture}(230,200)
\put(0,-5){\epsfxsize=230pt\epsffile[9 147 584 591 ]{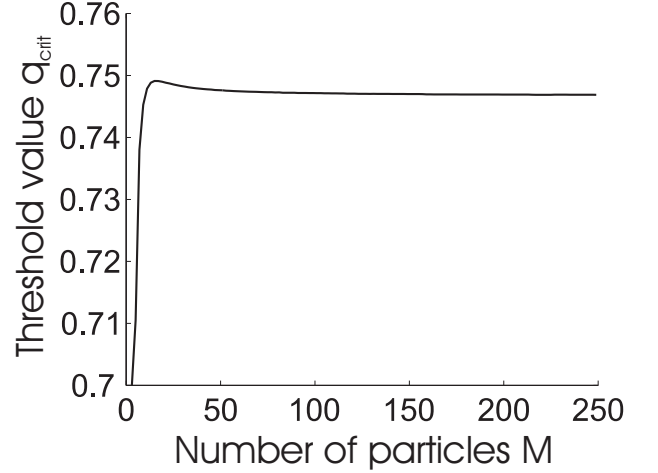}}
\end{picture}
\caption[]{Threshold value $q_{\rm crit}$ for imperfect local operations as a function of number of particles $M=N/2$ for $M$ odd.}
\label{Fig_thresholdp_anal}
\end{figure}

\subsection{Other graph states}

We have also numerically investigated other graph states and determined the corresponding threshold value. Here we have not only considered a single purification step as in the previous subsections, but analyzed the convergence of the whole purification procedure. In addition to bit flip errors in set $V_B$, we have also considered phase flip errors in set $V_A$ here. This error model has still the property that states belonging to the family $\rho_{\cal A}$ (Eq. (\ref{rhoA})) remain within this family throughout the procedure and the purification protocol $P1$ alone is sufficient to achieve purification.

For instance, we considered 2D cluster states corresponding to 2D lattices of different size. Note that a closed cluster state corresponds to periodic boundary conditions, while in an open cluster state the qubits at the border have fewer neighbors. We have investigated 2D cluster states which are closed in $x$ direction but open on $y$ direction on lattices of size $4 \times 3$ and $6 \times 3$ and found threshold values $p_{\rm min}^{(4\times 3)}=0.764$ and $p_{\rm min}^{(6\times 3)}=0.758$. For open 2D cluster states with $4 \times 4$ and $5 \times 3$ we find $p_{\rm min}^{(4\times 4)}=0.764$ and $p_{\rm min}^{(5\times 3)}=0.778$,  while for a completely closed $4 \times 4$ cluster state we have $p_{\rm min}^{(4\times 4)}=0.768$. 


We have also considered families of graph states $G_{(N,k)}$ specified by 2 parameters $N$ and $k$, where the number of vertices is given by $2N$ and $k$ specifies the degree of the graph. The set $V_A$ is given by all odd vertices $1,3,\ldots, 2N-1$, while the set $V_B$ consists of all even vertices $2,4,\ldots ,2N$. The edges of the graph are given by $\{j,j+1\},\{j,j+3\},\ldots\{j,j+2k-1\} \forall j$ odd and the addition is understood modulo $N$. That is, each vertex in $V_A$ is connected to the next $k$ vertices in $V_B$. The graph is translational invariant and has degree $k$.

We find that the threshold value is largely independent of both $N$ and $k$. For instance, we have for $G_{(10,3)},G_{(10,4)},G_{(10,5)},G_{(10,10)}$ that $p_{\rm min}=0.762$. Altogether, in the investigated regime $3\leq N \leq 10$, $2\leq k \leq N$ we find that the threshold value varies only between $0.768 \leq p_{\rm min} \leq 0.772$.


\section{Bipartite vs. Multipartite entanglement purification protocols}\label{comparebipartite}

In this section, we compare direct multiparticle entanglement purification protocols with protocols based on bipartite entanglement purification. For a large class of states we show: (i) In the case of perfect local operations, {\em any} protocol based on bipartite entanglement purification is less efficient ---in terms of the yield--- than a certain direct multiparticle entanglement purification protocol; (ii) In the presence of imperfect local operations, direct multiparticle entanglement purification protocols can perform better than protocols based on bipartite entanglement purification. That is, a wider range of states can be purified and the achievable fidelity of multipartite protocols is higher than with methods based on best known \cite{As04} bipartite entanglement purification protocols combined with teleportation.
While (i) justifies and motivates the investigation of multiparticle entanglement purification protocols from a principal point of view, (ii) makes thses protocols also interesting from a practical point of view.

In principle, bipartite entanglement purification seems to be sufficient to purify also multipartite entangled states. For instance, the following method accomplishes the desired task: all but two particles of a (noisy) multiparticle entangled state are measured and the resulting (noisy) bipartite entangled state is purified, thereby creating an (highly) entangled pair shared between two parties. This procedure is applied to several such pairs of parties, and the resulting pairs of highly entangled states can be used (e.g. by means of teleportation) to generate the desired multiparticle entangled state with high fidelity. However, as we shall see below, such a procedure may be quite inefficient and it is not obvious that all multipartite entangled states which can be purified by direct multipartite entanglement purification are also purificable using the procedure sketched above.

\subsection{Noiseless local operations}

In this section we compare the efficiency of direct multiparticle entanglement purification protocols with methods based on bipartite entanglement purification. In Ref. \cite{Mu99}, it was shown that in a restricted (but rather natural) scenario, where bipartite entanglement purification is combined with teleportation, direct multiparticle entanglement purification is more efficient for purifying $N$--particle GHZ states. In the scenario considered in Ref. \cite{Mu99}, $N-2$ particles of a single copy of a $N$--particle entangled mixed state are measured and the resulting bipartite entangled mixed state is purified by means of a bipartite recurrence protocol. Highly entangled pairs of particles shared between different pairs of parties created in this way are then used to generate ---by means of teleportation--- the desired $N$--particle GHZ state. To be specific, pairs between party 1 and $k$, $k\in\{2,3,\ldots,N\}$ are generated and a GHZ state is e.g. created by teleporting $N-1$ particles of a $N$--particle GHZ state, generated {\em locally} by party 1, to the remaining $N-1$ parties. The average number of copies of the initial state $\rho$ that are required to generate GHZ states with a certain fidelity turns out to be smaller for direct multiparticle entanglement purification, thereby indicating that such protocols can be more efficient than methods based on bipartite entanglement purification.  

However, the scenario considered by Murao et al. in Ref. \cite{Mu99} is a restricted one. For instance, it is assumed that bipartite entangled states are generated from a single copy of the initial multiparticle state $\rho$, and only a single copy of a $N$--particle GHZ state is generated from the produced bipartite entangled pairs using a specific procedure based on teleportation. Furthermore, only a specific bipartite entanglement purification protocol is considered. 
We will now show that for a large class of states, indeed {\em any} method which is at some point based on bipartite entanglement purification is less efficient than direct multiparticle entanglement purification, e.g. using multipartite generalizations of hashing or breeding. We emphasize that we do not specify the method how bipartite entanglement purification is employed, nor do we restrict ourselves to a specific way of combining the resulting bipartite entangled pairs to obtain the desired (purified) multiparticle entangled state. 

To this aim, we consider the most general method to purify multipartite entangled states which is based on bipartite entanglement purification. The only assumption is that at some point some kind of bipartite entanglement purification is used and thus maximally entangled pairs shared between pairs of parties are generated. These pairs are then used to generate (possibly several copies) of the desired multiparticle entangled state. We allow for joint manipulation of an arbitrary number of copies of the state at any point of the procedure, and for the most general bipartite entanglement purification protocol. Using the asymptotic inequivalence of multiparticle GHZ states and singlets, this is already sufficient to show that such protocols can be less efficient than e.g. multipartite breeding or hashing. 

We start with $M$ copies of a $N$--party entangled state $\rho$, $\rho^{\otimes M}$, which are manipulated by means of local operations and classical communication. This procedure involves bipartite entanglement purification and thus results in the generation of $m_{kl}$ copies of maximally entangled pairs in the singlet state $|\Psi^-\rangle_{kl}$ shared between parties $k$ and $l$. With help of another sequence of local operations assisted by classical communication these pairs are then transformed into $\tilde M$ copies of the desired multiparticle entangled state $|\chi\rangle$. 
The total procedure can be summarized as follows: 
\be
\rho^{\otimes M} \rightarrow \bigotimes_{k<l} |\Psi^-\rangle_{kl} \langle\Psi^-|^{\otimes m_{kl}} \rightarrow |\chi\rangle\langle\chi|^{\otimes \tilde M}.\label{process}
\ee
The yield of this procedure is given by $\tilde M/M$. In the following, we consider tripartite systems $N=3$ and analyze the special case where the input state $\rho$ is pure and in fact identical to the desired output state. That is, we consider $\rho=|\chi\rangle\langle\chi|$ where $|\chi\rangle$ is a three--particle GHZ--state, i.e. $|\chi\rangle$ is local unitary equivalent  to $1/\sqrt{2}(|000\rangle+|111\rangle)$. 

We make use of the following facts which were used in Ref. \cite{Li99} to proof the irreversibility of entanglement transformation between singlets and GHZ states: (i) The entropy of the reduced density operator with respect to system $l$, $l=1,2,3$ can only decrease under local operations and classical communication; (ii) The average increase in the relative entropy of entanglement of the system $(2-3)$ is smaller or equal than the average decrease in the entanglement of system 1 with the joint system (2-3) for any local protocol \cite{Li99}. Note that (ii) is valid only for pure states \cite{Li99}. If we consider a density operator $\sigma_{123}\equiv|\Psi\rangle\langle\Psi|$ corresponding to a pure state which is transformed by an arbitrary local protocol to an ensemble $\{p_k,\tilde\sigma_{123}^{(k)}\}$ we have that (i)
\be
S(\sigma_1) \geq \sum_k p_k S(\tilde\sigma^{(k)}_1),\label{entropy}
\ee
where $S(\sigma_1)=-\tr(\sigma_1 \log_2\sigma_1)$ with the reduced density operator with respect to system 1, $\sigma_1\equiv \tr_{23}(\sigma_{123})$, and similar for entropies of reduced density operator with respect to system $2,3$, while (ii) reads
\be
\sum_k p_k E_r(\tilde\sigma^{(k)}_{23})-E_r(\sigma_{23}) \leq S(\sigma_1)-\sum_k p_k S(\tilde\sigma^{(k)}_1).\label{relentropy}
\ee   
In these formulas, $E_r(\sigma_{23})$ denotes the relative entropy of entanglement of the reduced density operator $\sigma_{23}\equiv \tr_1(\sigma_{123})$,
\be
E_r(\sigma_{23})=\min_{\rho_{23} {\rm sep}} S(\sigma_{23}||\rho_{23}),
\ee
where the minimum is taken over all separable density operators $\rho_{23}$ and 
\be
S(\sigma_{23}||\rho_{23}) \equiv \tr(\sigma_{23}\log_2\sigma_{23}) - \tr(\sigma_{23}\log_2\rho_{23}),
\ee
is the relative entropy of $\sigma_{23}$ with respect to a bipartite state $\rho_{23}$.
For $\sigma=|\chi\rangle\langle\chi|$ we have that $S(\sigma_1)=S(\sigma_2)=S(\sigma_3)=1$, $E_r(\sigma_{23})=0$ (since $\tr_1(|\chi\rangle\langle\chi|)$ is separable), while e.g. for $\sigma=|\Psi^-\rangle_{12}\langle\Psi^-|$ one finds $S(\sigma_1)=S(\sigma_2)=1, S(\sigma_3)=0$, $E_r(\sigma_{12})=1, E_r(\sigma_{13})=E_r(\sigma_{23})=0$ and similar for singlets shared between parties $k,l$. 

We apply now Eq. (\ref{entropy}) to the second part of the process (\ref{process}) and find $m_{12}+m_{13} \geq \tilde M$ and similar for other reduced density operators, i.e. $m_{12}+m_{23} \geq \tilde M$, $m_{13}+m_{23} \geq \tilde M$. Combining these inequalities we obtain
\be
\tilde M \leq 2/3 (m_{12}+m_{13}+m_{23}).\label{one}
\ee
When applying Eq. (\ref{relentropy}) to the first part of the process (\ref{process}) we obtain
$m_{23}\leq M-m_{12}-m_{13}$, or equivalently
\be
(m_{12}+m_{13}+m_{23}) \leq M.\label{two}
\ee 
Combining Eqs. (\ref{one}) and (\ref{two}) one finds
\be
\tilde M\leq 2/3 M.\label{GHZbound}
\ee 
That is, for input states which are pure GHZ--state, the yield of {\it any} procedure based on bipartite entanglement purification to obtain again GHZ states is less or equal than $2/3$. This quantifies the amount of irreversability in the transformation of GHZ states to singlets and back. Clearly, the multipartite entanglement purification protocol ---which in this case consists of doing nothing--- has yield one. This already shows that for a certain input state, direct multiparticle entanglement purification is more efficient than any method based on bipartite entanglement purification. One can however easily prove a similar statement for a large class of input states. 

Consider the class of mixed states $\rho$ which can be obtained from GHZ--states $|\chi\rangle\langle\chi|$ by a deterministic local protocol, i.e. by a sequence of local operations and classical communication (LOCC). These states include, for instance, density operators of the form
\be
\rho(F)=F|\chi\rangle\langle\chi|+(1-F)\sigma,
\ee
where $\sigma$ is either an arbitrary separable density operator (e.g. $\frac{1}{8}\eins$) or any (classical) mixture of GHZ--states.

On the one hand, we have that for all such states the yield of any procedure based on bipartite entanglement purification to obtain GHZ states is less or equal than $2/3$. One can easily prove this by contradiction. Assume that a such a procedure, ${\cal M}$, with yield larger than $2/3$ would exist. In this case, one could first transform initial pure GHZ states in a deterministic way by LOCC to the state $\rho$ and apply ${\cal M}$ afterwards, thereby obtaining a yield for the conversion of GHZ states to GHZ states by a protocol based on bipartite entanglement purification larger than $2/3$. This clearly contradicts Eq. (\ref{GHZbound}), so such a procedure is impossible. 

On the other hand, we have that a multiparticle entanglement purification protocol exists which allows one to purify states of the form $\rho(F)$ with high yield, given $F$ is sufficiently large. In particular, a procedure consisting of depolarization of $\rho(F)$ to a GHZ--diagonal state (see Sec. \ref{depolarization}) leads to a state where the hashing protocol introduced in Ref. \cite{Sm00} (also discussed in Sec. \ref{hashing}) can be successfully applied. The yield of this protocol exceeds $2/3$ for a wide range of $F$, in fact approaches one for $F\rightarrow1$. That is, for a large class of input states, direct multiparticle purification is more efficient than any protocol based on bipartite entanglement purification.


\subsection{Imperfect local operations} 

It is also interesting to compare multiparticle entanglement purification protocols with protocols based on bipartite entanglement purification under realistic conditions, i.e. in the case where also local operations performed to manipulate entangled states are imperfect and give rise to errors. While above argumentation regarding the yield is based on the (idealized) assumption of perfect manipulation of an arbitrary large number of copies of a given state ---and the analysis is performed in full generality---, we will be concerned with practically implementable protocols in this section. That is, we consider entanglement purification protocols which operate in each round of the protocol only on a restricted number of copies of the state. We remark here that in the presence of imperfect local operations, protocols which operate on a large number of states simultaneously are very sensitive to errors in local operations, and therefore may become impractical anyway. The fact that imperfect local operations are involved in the purification procedure necessarily implies that no maximally entangled pure states can be created by any entanglement purification protocol and the corresponding yield ---defined as the average number of maximally entangled pure states produced per copy of $\rho$--- is zero. This suggests to use an adopted definition of the yield, e.g. to accept all output states which have a fidelity larger than some threshold value $F_0$. As we are only concerned with recurrence like entanglement purification protocols throughout this section ---which produce only a single copy of a state as output--- one can directly use the fidelity of this output state as criterion whether the protocol has created the desired state or not. The yield is then defined as the average number of produced states $\rho_k$ per copy of $\rho$ with fidelity larger than $F_0$, i.e. $F_k\equiv \langle\chi|\rho|\chi\rangle \geq F_0$, where $|\chi\rangle$ is the desired (pure) output state. Note that when considering general entanglement purification protocols, such a definition might not be adequate as several copies of output states might be entangled themselves. Such a definition implies that for $F_0\geq F_{\rm fix}$, i.e. the desired output fidelity is larger than the fixed point of the protocol, the protocol will have yield 0.       

We compare the recurrence protocol for multiparticle entanglement purification discussed in Sec. \ref{recurrence} with a scheme based on the bipartite entanglement purification protocol introduced in Ref. \cite{De96}). In the latter case, the protocol of Ref. \cite{De96} is first used to create bipartite entangled states, which are then used to create a multiparticle entangled state by some means, e.g. by teleportation. As we are interested only in the properties of the entanglement purification protocol, we have not specified the means how bipartite entangled states are combined to create a multiparticle entangled state. We have rather conservatively assumed that this process ---although it necessarily involves joint local operations on two qubits which may again be imperfect--- is error free, and the only source of errors results from the fact that no maximally entangled bipartite states can be created in the case of imperfect local operations. The achievable fidelity of the states is specified by the fixed point of the purification protocol, and is thus independent of the input state. That is, our analysis is valid for all (distillable) input states under this protocol. Note that the protocol of Ref. \cite{De96} is the up to now best known bipartite entanglement purification protocol with respect to the maximal reachable fidelity for a given noise level of imperfect local operations.  

For instance, if GHZ states with $N=3$ particles should be created, this involves at least two bipartite entangled states, e.g. shared between parties $A$ and $B$ [$A$ and $C$] respectively. The mixed state $\rho_{AB}$ corresponding to the fixed point of the bipartite entanglement purification of Ref. \cite{De96} is diagonal in the Bell--basis and can be described by ${\cal M}_B(|\Phi^+\rangle_{AB}\langle \Phi^+|)$ with $|\Phi^+\rangle=1/\sqrt{2}(|00\rangle+|11\rangle)$, where ${\cal M}_B$ is a map acting on $B$ only. A similar description exists for $\rho_{AC}$ in terms of a map ${\cal M}_C$ acting on $C$ only. The optimal case is that local operations in $A$ introduce no further errors and create out of two maximally entangled bipartite states a GHZ state. Since ${\cal M}_B, {\cal M}_C$ commute with all operations performed at $A$, the fidelity of the resulting state is upper bounded by the fidelity of the state ${\cal M}_B\circ {\cal M}_C (|GHZ\rangle_{ABC}\langle GHZ|)$. We have compared the maximal reachable fidelity $F_{\rm max}^{\rm MP}$ for our multiparticle entanglement purification with the upper bound for the method based on bipartite entanglement purification described above and observed that $F_{\rm max}^{\rm MP}$ is considerable larger as can be seen in Fig. \ref{Fig_comparebipartite}. This implies that under realistic conditions, i.e. when considering imperfect local operations, direct multiparticle entanglement purification schemes are advanteous as compared to schemes based on bipartite entanglement purification. In particular, if the given goal is to produce multiparticle entangled states with a given fidelity, this can be achievable using multiparticle purification, while the scheme based on bipartite purification fails to perform this task. That is, the yield of the multipartite protocol is nonzero, while the yield of the scheme based on bipartite entanglement purification is zero. Note that also in regimes where both schemes have non--zero yield, direct multipartite purification performs better than the scheme based on bipartite purification \cite{Mu99}.

If one considers the restricted scenario where a single copy of a multiparticle mixed state is manipulated to create bipartite states by means of measurements performed on the remaining particle, it might also happen that the bipartite state created in such a way is no longer (distillable) entangled, although the initial multiparticle state can be distilled by the multiparticle recurrence protocol \cite{Mu99}. That is, for these input states the yield for any such scheme based on bipartite entanglement purification is zero, while the multipartite entanglement purification protocol has non--zero yield. This is e.g. the case for three--qubit input states of the form 
\be
\rho(x)=x|GHZ\rangle\langle GHZ|+(1-x)/8\eins
\ee
with $ 1/5 \leq x\leq 1/3$. Any measurement performed by one of the parties on the state $\rho(x)$ produces a bipartite state of the form $\sigma(x)=x'|\Phi\rangle\langle\Phi|+(1-x')/4\eins$ with $x'=x$. It can easily be checked that $\sigma(x)$ is separable for $x\leq1/3$, while $\rho(x)$ is (distillable) entangled for $x>1/5$ if one allows for multiparticle entanglement purification. That is, the minimal required fidelity such that a (restricted) scheme based on bipartite purification can be successfully applied is larger than the one for schemes based on multipartite entanglement purification.

\begin{figure}[ht]
\begin{picture}(230,220)
\put(0,-5){\epsfxsize=230pt\epsffile[14 169 568 621]{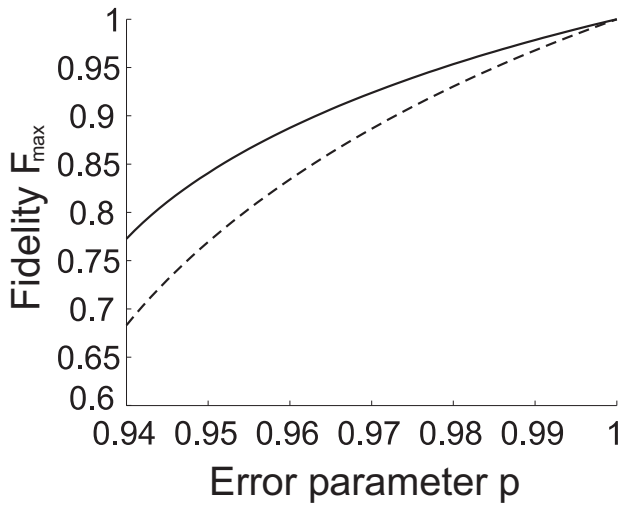}}
\end{picture}
\caption[]{Achievable fidelity of a linear cluster state with $N=4$ using direct multiparty entanglement purification (solid line) and conservative upper bound for methods based on bipartite entanglement purification (dashed line) for different errors in local operations $p$.}
\label{Fig_comparebipartite}
\end{figure}


\section{Private multiparticle entanglement}\label{private}



As we have seen in the previous section, it is not possible to distill
perfect cluster states using noisy apparatus. For bipartite protocols,
however, it was shown in Ref. \cite{As99} that even
using noisy apparatus it is possible to distill (asymptotically)
\emph{private} Bell pairs, i.\,e.\ Bell pairs which are only entangled
with the apparatus (i.e. the ``laboratories'') of the communication parties, but not with any
other degree of freedom. In a cryptographic scenario, this means that
the states of the pairs of particles are actively disentangled from any eavesdropper who
has, in the worst case, created the pairs, allowing her in principle 
to entangle them with additional degrees of freedom which he or
she controls.

In this section, we show that this is also possible with the cluster
purification protocol: if the parties only have imperfect apparatus
which they use to  purify cluster states, they will not be able to
create perfect cluster states; however, the final state will be
disentangled from all channel degrees of freedom. 

The proof is analogous to the proof of
Ref. \cite{As99}. In a first step, the noise which the
apparatus introduces during the purification process is replaced by a
simple toy-model, the \emph{lab demon}. The lab demon corresponds to an
intelligent source of noise, which uses a classical random number
generator in order to apply spin- and phase-flip operations on qubits,
according to a given probability distribution $f_{\mu\nu}$. The
action of the lab demon is thus the average of the ``flipped'' quantum
states:
\begin{equation}
  \label{eq:demon_noise}
  \rho_{ab\ldots} \rightarrow \rho_{ab}' = 
  \sum_{\mu\nu} \sigma_\mu^{(a)} \sigma_\nu^{(b)} \rho_{ab\ldots} 
  \sigma_\mu^{(a)} \sigma_\nu^{(b)}
\end{equation}
Here, $\rho_{ab\ldots}$ is a density operator of a quantum system,
which includes two qubits $a$ and $b$ which are located at one
specific party; however, it will include other qubits.
The lab demon acts on the two
qubits at the same time, since the quantum operations in the
purification protocols are two qubit operations; for that reason it
would be an over-simplification if we assumed that the noise acting on
two qubits is uncorrelated.

The labs demon keep notes on which Pauli operators were applied to
which qubits in which step of the purification process. As we will
show, the mere knowledge of this list will, in the asymptotic limit,
suffice to perfectly predict the state of the purified quantum
systems. In other words, from the lab demon's point of view, all
purified quantum systems end up in a pure state. Note that it is not
\emph{a priori} clear that the lab demon's knowledge would suffice for
the prediction, since the protocol includes measurements, and by
introducing errors, the measurement outcomes will be changed, possibly
leading to different choices by communicating parties, who might throw
away qubits which they should have kept and \emph{vice versa}.

From the list of errors, the lab demons calculate the so-called
\emph{error flags}. An error flag as a piece of classical information,
which is ``attached'' to each copy of the cluster state. In case of a $n$ qubit
cluster state, we need $n$ classical bits $\vec\lambda^{(j)} =
(\lambda_1^{(j)},\ldots \lambda_n^{(j)}) \in \{0,1\}^n$ for the
error flag. Here, the index $j$ denotes the number of the cluster
state in the ensemble of all cluster states.  Initially, before the
first step of the purification process, all error flags are set to
zero, i.e. $\vec \lambda^{(j)} = (0,\ldots,0)$ for all $j$.
Whenever the $i$-th lab demon applies a phase flip operation
($\sigma_z$) to the $i$-th qubit of cluster state $j$, in the error
flag $j$ the $i$th bit is flipped, i.e.
 \bea
  \label{eq:cluster_flag_flip}
  \vec\lambda^{(j)} &=& (\lambda_1^{(j)},\ldots \lambda_i^{(j)},\ldots
,\lambda_n^{(j)}) \nonumber \\ 
&\rightarrow&
\vec\lambda'^{(j)} = (\lambda_1^{(j)},\ldots 
\bar\lambda_i^{(j)},\ldots \lambda_n^{(j)}).
\eea
If he
applied an amplitude flip operation ($\sigma_x$), the
\emph{adjacent} bits of the error flag (associated with the neighbors of qubit $i$ in the cluster) are flipped, i.e.
\bea
  \label{eq:cluster_flag_flip_2}
  &&\vec\lambda^{(j)} = (\lambda_1^{(j)},\ldots \lambda_{i-1}^{(j)}
  \lambda_i^{(j)}\lambda_{i+1}^{(j)},\ldots 
  ,\lambda_n^{(j)}) \nonumber 
  \\ &\rightarrow&
  \vec\lambda'^{(j)} = (\lambda_1^{(j)},\ldots 
  \bar\lambda_{i-1}^{(j)},
  \lambda_i^{(j)},\bar\lambda_{i+1}^{(j)},\ldots \lambda_n^{(j)}).
\eea


In both purification sub-protocols $P1$ and $P2$, two
cluster states are combined, one of which (probabilistically)
survives. The error flag vector of the remaining state is then given
by a function of the both error flags of the input cluster states.
This function is called the \emph{flag update function}\index{flag
  update function!cluster purification protocol} for protocol
$P1$ and $P2$, respectively.

\subsection{The flag update function}

The error flags of the first and second cluster state are given by the
vectors $(\kappa_1, \kappa_2, \ldots \kappa_n)$ and $(\lambda_1,
\lambda_2, \ldots \lambda_n)$, respectively.  For the sub-protocol
$P1$, the flaf update function maps these $2n$
classical bits onto $n$ classical bits, i.e. \[f_\mathrm{flup}:
\{0,1\}^{2n} \rightarrow \{0,1\}^{n},\] with
\bea
  \label{eq:cluster_flag_update_function}
      &(\kappa_1, \ldots \kappa_n, \lambda_1,\ldots \lambda_n)
      \\\nonumber 
      & \quad \mapsto \left\{ 
        \begin{array}{r l}
          (\kappa_1\oplus\lambda_1, \kappa_2, \kappa_3\oplus\lambda_3,
          \kappa_4, \ldots ) & 
          \mbox{if $\kappa_{2k} \oplus \lambda_{2k} = 0 \forall k$} \\
          (0,0, \ldots, 0) & \mathrm{otherwise}
        \end{array}
      \right.
\eea
The first line of the definition takes into account how errors are
propagated through the CNOTs operation. This means, that having
applied a certain pattern of error operations (given by the error flag
vectors) \emph{before} the CNOTs operation is equivalent to applying a
different pattern of error operations (given by the new error flag
vector, $\vec\kappa' = f_\mathrm{flup}(\vec\kappa, \vec\lambda)$)
\emph{after} the CNOTs operation. The second line in the definition is
the so-called \emph{reset rule} (see \cite{As04}).

It is necessary to introduce the reset rule, otherwise the security
proof does not work. The reset rule is found by the following
heuristics, which is equivalent to the heuristics used for the
bipartite protocol:

The flag update function is only used if in the protocol the first
cluster state is kept. This is the case if the values of all even
eigenvalues of the second cluster state are equal to zero, i.e.
$\mu_2\oplus\nu_2 = \mu_4\oplus\nu_4 = \ldots = 0$. If this is the case, and, at the same time, at least one of
the ``new'' error flags associated with the even qubits of the second
cluster state, has the value ``1'', then the errors in the history of
the protocol have summed up in such a way that the first cluster state
is kept. This is the case even though it would have been discarded if there had not
been introduced any errors. In that case, the error flag of the
remaining cluster state is set (re-set) to $(0,0,\ldots 0)$. Note
that this coincidence of the two before-mentioned conditions happens
infrequently; in fact, in the course of the purification process, the
probability for this coincidence converges to zero.

For the sub-protocol $P2$, the flag update function can be
constructed by exchanging even and odd numbers. Using this method, an
error flag can be calculated for each cluster state in each step of
the purification process. By construction, the error flags only depend
on the errors introduced by the lab demons. 

\subsection{The conditional fidelity}

Using the error flag of each cluster state, it is now possible to
divide the ensemble of all cluster states into $2^n$ sub-ensembles.
The state of the sub-ensemble, which belongs to the error flag
$\vec\lambda$, is labelled $\rho^{(\vec\lambda)}$. It is convenient
to normalize the density operators of the sub-ensembles to the
relative frequency of the respective error flags, so that the
(normalized) total density operator is just the sum of the
density operators of the sub-ensembles. Using this convention, we
define the \emph{conditional fidelity} 
\begin{equation}
  \label{eq:conditional_fidelity}
  F^\mathrm{cond} = \sum_{\vec\lambda} \langle\Psi_{\vec\lambda}|
  \rho^{(\vec\lambda)} |\Psi_{\vec\lambda}\rangle;
\end{equation}
here, the state $|\Psi_{\vec\lambda}\rangle = \Ket{\Psi_{\lambda_1,\ldots,
  \lambda_n}}$ denotes the cluster state. The conditional fidelity is a measure
for the \emph{purity} of the cluster states from the lab demons point
of view: since the lab demons know the error flags of all cluster
states, they can use this information to transform the ensemble of all
cluster states into an ensemble with fidelity $F^\mathrm{cond}$. In
contrast, the usual fidelity, which is just the overlap of the total
density operator with the cluster state $\Ket{\Psi_{\bm 1}}$, is
given by $F = \Bra{\Psi_{\bm 0}}\rho_\mathrm{total}
\Ket{\Psi_{\bm 0}} \equiv \langle\Psi_{\bm 0}|\sum_{\vec \lambda} \rho^{(\vec \lambda)}|\Psi_{\bm 0}\rangle$. 

In order to investigate the behavior of the conditional fidelity in
the course of the purification process, it is necessary to calculate
the states of all $2^n$ sub-ensembles in each step of the purification
process. Again, it is useful to note that all sub-ensembles are
diagonal in the cluster basis; the states of all sub-ensembles is thus
given by a real $2^n \times 2^n$-matrix $M$. The columns of this
matrix are the vectors of the diagonal elements of the density
matrices describing the sub-ensembles. Using this convention, physical
action on the qubits is described by a matrix multiplication from the
left, and a modification of the error flags is described by a matrix
multiplication from the right. 

Applying a one-qubit depolarizing channel is thus formally equivalent
to a super-operator acting on the matrix of the diagonal vectors. To
be specific, an error operation on qubit $i$ results in flips of the
cluster bit $i-1$, $i$, or $i+1$, respectively (see Sec. \ref{properties}). Simultaneously, bit $i-1$, $i$, or
$i+1$, of the error flag is flipped (Eq.~\ref{eq:cluster_flag_flip}
and \ref{eq:cluster_flag_flip_2}). The result of applying the error
operator $\sigma_\nu^{(i)}$ is thus (for $\nu=x,y,z$)
\bea
  \label{eq:cluster_flipped_matrices}
    M_z^{(i)} &= \tilde\sigma_x^{(i)} M \tilde\sigma_x^{(i)} \\
    M_x^{(i)} &= \tilde\sigma_x^{(i-1)} \tilde\sigma_x^{(i+1)} 
    M \tilde\sigma_x^{(i-1)} \tilde\sigma_x^{(i+1)}\\
    M_y^{(i)} &= \tilde\sigma_x^{(i-1)} \tilde\sigma_x^{(i)}
    \tilde\sigma_x^{(i+1)} M \tilde\sigma_x^{(i-1)} 
    \tilde\sigma_x^{(i)} \tilde\sigma_x^{(i+1)}.
\eea
Here, \(\tilde\sigma_x^{(i)}\) is the $i$-th cluster bit flip
operator, which looks in the cluster basis like the Pauli operator
$\sigma_x$ in the computational basis.
Under the action of the depolarizing channel on qubit $i$, the matrix
$M$ is thus transformed into a convex combination of matrices
$M_z^{(i)}$,
\begin{equation}
  \label{eq:cluster_convex_sum_matrices}
  M \rightarrow f_0 M + \sum_{\nu=1,2,3} f_\nu M_\nu^{(i)}.
\end{equation}


The application of the CNOT operations and the following measurement
can be implemented by the following algorithm. $M$ is the matrix of
the diagonal elements of the sub-density-matrices before the
sub-protocol $P1$ is applied, and $M'$ is the result matrix.
The algorithm calculates for all combinations of cluster states the
results of the CNOT operations. 
We check the result of the measurement of cluster state 2; if the
results are such that the first cluster state is kept, we calculate
its state $\ket{\Psi_{\vec k'}}$, and performs for all combinations of
error flags the following steps: (i) calculate the value of the new
error flag $\vec \lambda'$, using the flag update function,
(ii) add to the matrix element ${M_{\vec k'}^{\vec \lambda'}}'$ the
joint probability that cluster state one was in the state $\ket{\Psi_{\vec
  k}}$ with error flag $\vec \kappa$ \emph{and} that the cluster
state two was in the state $\ket{\Psi_{\vec l}}$ with error flag $\vec
\lambda$.
The result of this algorithm is the new matrix $M'$, which contains
the (non-normalized) states of all sub-ensembles after one step in the
purification process.

For the sub-protocol $P2$, a similar algorithm can be given. 
As a result, we find that the conditional fidelity converges to unity
in the course of the protocol, while the usual fidelity converges to
some value $F^\mathrm{max}$ (see
Fig.~\ref{fig:conditional_fidelity}).

\begin{figure}[ht]
\begin{picture}(230,170)
\put(-5,0){\epsfxsize=230pt\epsffile[125 447 483 671]{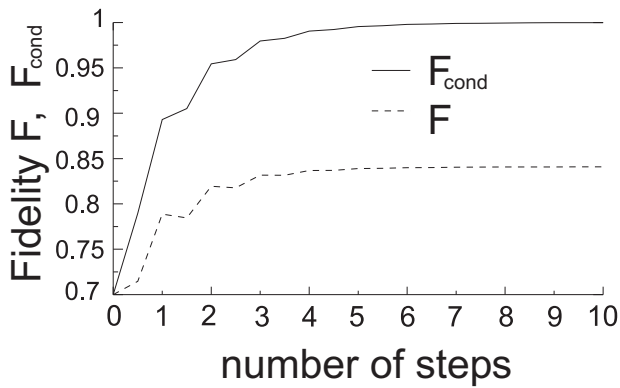}}
\end{picture}
\caption[Fidelities {$F$ and $F^\mathrm{cond}$} in the cluster
  purification protocol]{The fidelity and the conditional fidelity as a function of the number of steps in the purification protocol.}
\label{fig:conditional_fidelity}
\end{figure}



\section{Applications}\label{applications}

In this section, we discuss some possible application of our multiparticle entanglement purification protocols. 
Given the fact that the produced entanglement is private, one may be able to use multiparty entangled states produced in this way for secure communication and computation, e.g. secret sharing or secure function evaluation. However, a careful analysis of the protocol in the presence of a number of distrustful parties is required before a final conclusion can be drawn. 

\subsection{Purification of concatenated error correcting CSS codes}

A more direct application of the protocol is in the context of quantum error correction. There exist quantum error correction codes which correspond to graph states. In particular, Schlingemann and Werner \cite{Sc01} have shown that for certain graph states coding into an error correcting code can be achieved via a single (Bell) measurement. That is, a certain graph state $|\Psi\rangle_{\cal G}$ serves as ``encoding state''  and an unknown state $|\varphi\rangle =\alpha |0\rangle +\beta |1\rangle$ (which contains the quantum information which should be encoded) can be encoded by performing ``teleportation'', where $|\Psi\rangle$ plays the role of the channel (singlet) in the original teleportation scheme. The result of this procedure is an encoded state $\alpha|0\rangle_L + \beta|1\rangle_L$, where the codewords $|0\rangle_L,|1\rangle_L$ are two orthogonal graph states corresponding to the same graph $\tilde{\cal G}$ which is directly related to the original graph ${\cal G}$. We remark that $|\Psi\rangle_{\cal G}$ completely determines the kind of encoding, in particular the properties of the corresponding error correcting code. In particular, $|\Psi\rangle_{\cal G}$ can be chosen in such a way that it corresponds to a concatenated code with several concatenation levels.  

The basic idea here is to use multiparty entanglement purification to purify the encoding states $|\Psi\rangle_{\cal G}$. That is, the resource for encoding is purified and then used to encode the desired quantum informational. We emphasize that independent of the kind of code used (in particular, independent of the number of concatenation levels when using a concatenated code), the final encoding takes place by performing a {\em single} Bell measurement. That is, a measurement in the basis $\{|\Phi_i\rangle\}$ with $|\Phi_i\rangle = \eins\otimes\sigma_i |\Phi^+\rangle$. As in the original teleportation scheme, one can perform local unitary operations depending on the measurement outcome such that the resulting state is for all possible measurement outcomes given by $\alpha|0\rangle_L + \beta|1\rangle_L$.

Many of graphs corresponding to error correcting codes are two--colorable which ensures that our entanglement purification protocol can be successfully applied. In particular, all CSS--codes are equivalent to two--colorable graph states \cite{Rai04}. For instance, the graph corresponding to the seven qubit steane code (a CSS $(7,1,3)$ code) is given by a cube (see Fig. \ref{cube}), which is clearly two-colorable. Note that also the concatenated code of this kind may correspond to a two--colorable graph state. In fact, the corresponding graph at the next concatenation level can be obtained by appending to each vertex of the cube another cube with seven new vertices and measuring the vertices of the initial cube in the eigenbasis of $\sigma_x$. By concatenating this procedure, i.e. appending new cubes on each of the vertices and performing the corresponding measurement, one obtains the graph corresponding to the encoding states for concatenated CSS code. When postponing the $\sigma_x$ measurements, we have in fact that the resulting graph state is still two-colorable. Note that the measurement implements the encoding procedure, i.e. information which is initially represented in the state of the qubit of a single vertex is encoded into the qubits of seven new vertices.

We find that the entire encoding circuit which serves to encode a given qubit into a certain (concatenated) code of a larger number of qubits can be replaced by the following simple procedure. One first creates the graph state corresponding to a cube, where each vertex of the cube may have another cube appended (and so on when dealing with more concatenation levels). Note that the vertices of new cubes which are appended are not yet measured. The qubit to be encoded is then measured together with the particle $V_1$ of the first cube in the Bell basis. A sequence of measurements in the eigenbasis of $\sigma_x$ completes the encoding procedure: one starts with the vertices of the cube at concatenation level one, followed by the vertices of the cubes at concatenation level 2 etc., until only qubits at the highest concatenation level are left. That is, the quantum information of the initial qubit (one logical bit) is now encoded into $7^k$ physical qubits, where $k$ gives the number of concatenation levels. In case all operations involved in this procedure are perfect, this results in an error free encoding. However, given that operations used in the manipulation and creation of the states are imperfect, the encoding will not be perfect. In particular, the main difficulty in the procedure described above is the creation of the multiparticle entangled graph state corresponding to the graph with (appended) cubes. Since this graph is two--colorable, one can apply our entanglement purification protocol to improve the fidelity of this state ---and hence improve the achievable fidelity of encoding.  


\subsection{Purification of algorithms}

We also note that graph states are an algorithmic resource. In the same way as a cluster state is a universal resource for measurement based quantum computation, certain graph states are a specific resource for a given quantum algorithm \cite{Ra03}. That is, a quantum algorithm (e.g. a quantum fourier transformation) can be implemented by consuming an algorithmic specific resource ---the graph state in question--- by performing local measurement only. Again, in the presence of imperfect operations the corresponding graph state may not be available with unit fidelity. However, our entanglement purification protocol allows one to increase the fidelity of the graph state and hence the fidelity of the implementation of the algorithm. This opens up new possibilities for the use of EPP in quantum computation \cite{Du03QC} and for fault tolerant computation \cite{Ra04}. Important issues in this context are fault tolerance and error correction, which will be discussed in more detail in a forthcoming publication \cite{Ra04}.


\section{Experimental realization}\label{experiment}

In this section, we propose an experimental realization of multiparticle entanglement purification protocols using neutral atoms trapped in optical lattices \cite{Ja98,Gr02,Gr02b,Ma03,Ma03b}. We show that multiparticle entanglement purification protocols can be used in such systems to increase the fidelity of cluster states. In particular, we consider the purification of 1D cluster states in a 2D lattice, which can be straightforwardly generalized to the purification of 2D cluster states in a 3D lattice. We show on the one hand that the effect of decoherence can be overcome by using a scheme based on {\em entanglement pumping}. On the other hand, we find that implementing the standard recurrence scheme allows one to increase the achievable fidelity of cluster states. This result is quite remarkable, as the same imperfect operations are involved in the creation of the cluster state and in the purification process. 

\subsection{Physical Implementation}\label{impl}

Consider a two--dimensional $N \times N$ optical lattice filled with one atom per lattice site. Internal states of the atoms ---which constitute the qubits--- can be manipulated by means of laser pulses. While in the present experimental setup addressing of individual atoms is still a problem, there are proposals to overcome this limitation, for example by expanding the lattice, or by using reloading techniques into lattices with larger spacing. In the following we will assume that individual addressing of the atoms is possible.
Interactions between neighboring atoms take place e.g. by state--selectively shifting the lattice, leading to a state dependent collisional phase arising from controlled cold collisions \cite{Ja98,Ma03b}. The interaction Hamiltonian describing a lattice shift in the $x$--direction is given by 
\be
H_x=4g(t) \sum_{(k,l)} (1-\sigma_z^{(k,l)})/2 \otimes (1-\sigma_z^{(k+1,l)})/2, 
\ee
where $(k,l)$ labels the $(x,y)$--coordinate of the atom. Note that for $\int g(t) dt=\pi$, such an interaction produces $N$ copies of one dimensional cluster states along the $x$--direction of the lattice when applied to states of the form $(|0\rangle+|1\rangle)^{\otimes N^2}$. These states can than be purified by using lattice shifts along the $y$--direction as follows. In a first step, we want to (simultaneously) implement protocol $P1$ to the linear cluster states in rows $2l$ and $2l+1$.  
We have that $H_y$ is equivalent up to local unitary operations to the Ising Hamiltonian 
\be
H_I=g(t) \sum_{(k,l)} \sigma_z^{(k,l)}\otimes \sigma_z^{(k,l+1)}. 
\ee
On the one hand, applying $H_I$ for $\int g(t) dt =\pi/2$, followed by the local unitary operation $\sigma_x$ applied to particles $(k,4l)$, $(k,4l+1)$ before and after another application of $H_I$ for $\int g(t) dt =\pi/2$ results into an effective interaction $\tilde H_I$ which performs phase gates between rows $2l$ and $2l+1$, while the interaction between rows $2l+1$ and $2l+2$ is cancelled. By means of local operations performed before and after the application of $\tilde H_I$, one can convert each of these phase gates into a CNOT gate with the freedom of choosing source and target for each pair of particles independently. This allows one to implement protocol $P1$ simultaneously to $N/2$ pairs of linear cluster states with a total of two sweeps of the lattice in $y$--direction. The final measurement of the cluster states in rows $4l-1$, $4l$ leaves us ---in the case the measurement was successful--- with linear cluster states of improved fidelity at rows $4l+1$, $4l+2$, which can further be purified by applying protocol $P2$ in a similar way. Note that iterations of the protocol may involve lattice shifts over longer distances. 

\subsection{Improved Fidelity}

We now analyze the purification protocol sketched above in the case where the operations involved in the procedure are imperfect. Specifically, we consider the interactions between neighboring atoms ---and thus also the resulting CNOT operations--- to be imperfect. There are various possible sources of imperfections, ranging from imperfection in the laser manipulation of the internal states of the atoms to fluctuations in the desired interaction time.  
We will consider a simple model to describe imperfections in the gates. As in the previous discussion, we describe imperfect operations by a completely positive map which consists of first applying a partially depolarizing channel with error parameter $p$ to the individual particles followed by the perfect operation (see discussion in Sec. \ref{ILO}, in particular Eq. (\ref{noisyU})). To be consistent, we assume that the {\em same} imperfect operations are involved in the creation of the cluster state and in the purification procedure. In the procedure sketched above, both processes, the creation of cluster states and the implementation of gates in the purification, are physically implemented by the same procedure and thus our assumption that both processes suffer from same imperfections is reasonable in such systems. In particular, cluster states are created by shifting the lattice along the $x$--direction, while interactions between neighboring atoms resulting in a CNOT operation (used for entanglement purification) are implemented by a lattice shift along the $y$--direction.

We now compare the fidelity of 1D cluster states created directly in the lattice by simply shifting it along the $x$--direction with the achievable fidelity when using above purification procedure. Up to local unitary operations, the gate operation involved in the creation of the cluster state is given by
\bea
U(t)&=&e^{-i t g(t) \sum_{(k,l)} \sigma_z^{(k,l)}\otimes\sigma_z^{(k+1,l)}}\nonumber\\
&=&\prod_{(k,l)} e^{-i t g(t) \sigma_z^{(k,l)}\otimes\sigma_z^{(k+1,l)}}, 
\eea
i.e. corresponds to a sequential application of phase gates to neighboring particles. Note that we have $\int g(t) dt=\pi/2$ in this case, and that initially all atoms are prepared in state $1/\sqrt{2}(|0\rangle+|1\rangle)$. Assuming that each of these phase gates is imperfect and modelled by Eq. (\ref{noisyU}), one readily obtains the fidelity of the resulting state. The results for $p=0.99$ for different sizes of the cluster state are summarized in Table \ref{tab}.

The maximal achievable fidelity $F_{\rm max}$ of the recurrence protocol implemented in an optical lattice when considering imperfect CNOT operations can be readily determined. We assume that the state created by the lattice shift along the $x$--direction is used as input state for the purification protocol. As one can see from Table \ref{tab}, the achievable fidelity can be significantly enhanced by the purification procedure, although the operations involved in the creation of the cluster state and in the purification procedure have the same fidelity.

\begin{table}
\begin{tabular}[t]{|c|c|c|} \hline
$N=2$&  $F=0.9900$& $F_{\rm max}=0.9889$ \\ \hline 
$N=3$&  $F=0.9753$& $F_{\rm max}=0.9836$\\ \hline
$N=4$&  $F=0.9608$& $F_{\rm max}=0.9785$\\ \hline
$N=5$&  $F=0.9465$& $F_{\rm max}=0.9734$\\ \hline
$N=6$&  $F=0.9324$& $F_{\rm max}=0.9681$\\ \hline
\end{tabular}
\caption[]{Fidelity $F$ of linear cluster state of size $N$ created using imperfect operations with error parameter $p=0.99$ and achievable fidelity $F_{\rm max}$ when using entanglement purification with noisy operations of the same quality.}
\label{tab}
\end{table}

\subsection{Entanglement pumping}

In the discussion of the entanglement purification protocol in the previous paragraph, we assumed that the original recurrence protocol is applied. In particular, this involves in each step of the protocol a manipulation of two identical copies of the state obtained in the preceding round of the protocol. A modified protocol which is called ``entanglement pumping'' operates always on one copy of the state to be purified (whose fidelity increases during the process) and on a second state of some standard form. The fidelity of the second state is always the same throughout the procedure. That is, the input state at stage $k$ of the protocol is given by $\rho=\rho_{k-1}\otimes\rho_0$, where $\rho_{k-1}$ is the state obtained in the previous round, while $\rho_0$ is the initial state. Note that also in this case, protocols $P1$ and $P2$ are iteratively applied. 

On the one hand, entanglement pumping offers the advantage to use always states of a certain standard form which may be easy to produce, e.g. they may arise from sending a locally prepared cluster states through noisy quantum channels to several parties. The possibility to produce these states on demand reduces the required storage capabilities of the whole procedure, as only two copies of the state have to be stored simultaneously when using entanglement pumping, while the application of the standard recurrence protocol typically requires simultaneous storage of hundreds of copies of the state. On the other hand, entanglement pumping has the disadvantage that even in the case of noiseless local operations no maximally entangled pure states can be produced. Iterative application of the protocol only allows one to increase the fidelity of the state by a certain amount. By applying a nested entanglement pumping scheme (introduced in Ref. \cite{Du03QC}) one can overcome this limitation. A few nesting levels ---which correspond to the number of extra copies of the state which need to be stored simultaneously--- typically suffice to reach fidelities close to those achievable with the standard recurrence protocol.   

It turns out that entanglement pumping ---in contrast to the standard recurrence scheme--- does not allow one to increase the fidelity of cluster states if the noisy operations used to create the states are also used in the purification procedure. However, entanglement pumping may still be used to {\em maintain} high fidelity cluster states in the presence of decoherence, i.e. to stabilize these states. In optical lattice systems the implementation of entanglement pumping is even simpler than the implementation of the standard recurrence scheme. The production of the linear cluster state $\rho_0$  can be accomplished by a lattice shift along the $x$ direction. The state to be purified should in this procedure not participate on the interaction. One possibility to achieve this is by transferring the state of the neutral atoms to internal states which are trapped in an independent lattice potential which is not moving. Another option is to apply two lattice shifts intercepted by local unitary operations on this copy of the state which are chosen in such a way that the interaction cancels. This is similar to the procedure described in Sec. \ref{impl} to implement CNOT gates between certain pairs of atoms, while no interaction takes place between certain other pairs. Note that a implementation of the entanglement pumping protocol for a $N$--particle linear cluster state only requires a $N \times 2$ lattice.



\section{Summary and conclusions}\label{summary}

In this paper we have analyzed in detail entanglement purification protocols (recurrence schemes and hashing protocols) which are capable of purifying arbitrary two--colorable graph states. For the recurrence schemes, we found that 
(i) The purification regime of the protocol for graph states does depend on the degree of the graph, but is independent of the number of particles $N$. That is, the resulting state $\rho$ arising from a perfect cluster state due to channel noise (local decoherence) can be successfully distilled using the protocol as far as the decoherence per particle is below a certain threshold value which depends on the degree of the graph, but is independent of $N$; 
(ii) In the case of {\em noisy local control operations}, we observe that the corresponding threshold for local control operations such that the protocol can be successfully applied is for cluster states (and similar states where the degree of the corresponding graph does not depend on $N$) is independent of the size of the system. In contrast, the requirements to purify GHZ states become more stringent for increasing $N$.
We have that (i) and (ii) together suggest that our protocol may be used for practical applications to purify certain states, e.g. in the context of purification of quantum algorithms or concatenated quantum error correction codes. We have also shown that the entanglement created by our purification protocol is private, an important feature for possible applications for secure communication and computation. We have compared multiparty entanglement purification protocols with protocols based on bipartite entanglement purification and found that direct multiparticle entanglement purification is not only more efficient, but also the achievable fidelity of the state is larger. Finally we proposed a possible experimental implementation of the protocol based on neutral atoms in an optical lattice. This scheme allows one to increase the fidelity of cluster states created in such systems.

We are confident that multiparticle entanglement purification will prove a useful tool in various branches of quantum information, ranging from measurement based quantum computation over quantum error correction to applications in quantum security and quantum communication.


{\em Note added:} After completion of this work, we have learned about similar results on a hashing method to purify CSS states by Kai Chen and Hoi-Kwong Lo \cite{Ch04}.

We thank R. Raussendorf, M. Grassl, M. Hein and P. Aliferis for discussions and remarks. 
This work was supported by the Deutsche Forschungsgemeinschaft and the European Union (IST-2001-38877,39227). W.D. acknowledges support from the \"Osterreichische Akademie der Wissenschaften through project APART.








\begin{thebibliography}{99}

\bibitem{Be93}
C. H. Bennett, G. Brassard, C. Cr\'{e}peau, R. Jozsa, A. Peres, and W. K. Wootters, Phys. Rev. Lett. {\bf 70}, 1895 (1993).

\bibitem{Be91}
C. H. Bennett and S. J. Wiesner, Phys. Rev. Lett. {\bf 69}, 2881 (1992).

\bibitem{Ek91}
A. Ekert, Phys. Rev. Lett. {\bf 67}, 661 (1991).

\bibitem{secure}
A. Karlsson, M. Koashi, and N. Imoto, Phys. Rev. A {\bf 59}, 162 (1999);
M. Hillery, V. Bu\v{z}ek, and A. Berthiaume, Phys. Rev. A {\bf 59}, 1829 (1999).
R. Cleve, D. Gottesman, and Hoi-Kwong Lo, Phys. Rev. Lett. {\bf 83}, 648 (1999).

\bibitem{Cr02}
C. Cr\'{e}peau, D. Gottesman and A. Smith, Proceedings of the thirty-fourth annual ACM symposium on Theory of computing,
Montreal, Quebec, Canada, 643 - 652 (2002) (E-print quant-ph/0206138).

\bibitem{Wi93}
D. J. Wineland, J. J. Bollinger, W. M. Itano, F. L. Moore, and D. J. Heinzen, Phys. Rev. A {\bf 46}, R6797 (1992).

\bibitem{Hu97}
S.F Huelga, C. Macchiavello, T. Pellizzari, A.K. Ekert, M. B. Plenio, J.I. Cirac, 
Phys. Rev. Lett. {\bf 79}, 3865 (1997).

\bibitem{Br01}
H.-J. Briegel and R. Raussendorf, Phys. Rev. Lett. {\bf 86}, 910 (2001). 

\bibitem{Ra98}
R. Raussendorf and H.-J. Briegel, Phys. Rev. Lett. {\bf 86}, 5188 (2001). 


\bibitem{Be96}
C. H. Bennett, G. Brassard, S. Popescu, B. Schumacher, J. A. Smolin and W. K. Wootters
Phys. Rev. Lett. {\bf 76}, 722 (1996);
C. H.Bennett, D. P. DiVincenzo, J. A. Smolin and W. K. Wootters, 
Phys. Rev. A {\bf 54}, 3824 (1996).

\bibitem{De96}
D. Deutsch, A. Ekert, C. Macchiavello, S. Popescu, and A. Sanpera, 
Phys. Rev. Lett. {\bf 77 }, 2818 (1996).

\bibitem{Mu99}
M. Murao,  M. B. Plenio, S. Popescu, V. Vedral, and P. L. Knight, Phys. Rev. A {\bf 57}, R4075 (1998).

\bibitem{Sm00}
E. N. Maneva and J. A. Smolin, in {\em Quantum Computation and Quantum Information}, edited by J. Samuel J. Lomonaco (American Mathemathical Society, Providence, RI, 2002), Vol. 305 of {\em AMS Contempurary Mathematics}; see also E-print quant-ph/0003099.

\bibitem{Du03}
W. D\"ur, H. Aschauer and H.-J. Briegel, Phys. Rev. Lett. {\bf 91}, 107903 (2003).

\bibitem{Ra03}
R. Raussendorf, D. Browne and H.-J. Briegel, Phys. Rev. A {\bf 68}, 022312 (2003).


\bibitem{Rai04}
E. Rains and Hoi-Kwong Lo, private communication (see also Ref. \cite{Ch04}).

\bibitem{BrEi00}
J. Eisert and H.-J. Briegel, Phys. Rev. A {\bf 64}, 022306 (2001).

\bibitem{Sc01}
D. Schlingemann and R. F. Werner, Phys. Rev. A {\bf 65},  012308 (2002);
M. Grassl {\em et al.}, in {\em Proceedings of the ISIT, Lausanne, 2002} (IEEE, Piscataway, NJ, 2002), p. 45.


\bibitem{He03}
M. Hein, J. Eisert and H.-J. Briegel, Phys. Rev. A {\bf 69}, 062311 (2004).

\bibitem{Ne03}
M. Van den Nest, J. Dehaene, and B. De Moor, Phys. Rev. A {\bf 69}, 022316 (2004);
M. Van den Nest, J. Dehaene, and B. De Moor, Phys. Rev. A {\bf 70}, 034302 (2004).


\bibitem{noteCNOT}
The CNOT operation is defined by $|i\rangle_A|j\rangle_B \rightarrow |i\rangle_A|i\oplus
j\rangle_B$, where $\oplus$ denotes addition modulo 2. 


\bibitem{footnotevertex}
States corresponding to graphs with a very high symmetry may not exactly follow this behaviour. Due to the fact that all neighbouring particles of a vertex are affected by a bit flip error occuring at this vertex, sometimes errors may annihilate each other or affect the whole system in such a way that only a certain subspace is populated. As low rank density matrices are sometimes quite robust against noise, this can result in a behaviour for certain graph states not expected by this intuitive explanation.   

\bibitem{Du03a}
W. D\"ur and H.-J. Briegel, Phys. Rev. Lett. {\bf 92}, 180403 (2004).


\bibitem{Li99}
N. Linden, S. Popescu, B. Schumacher and M. Westmoreland, quant-ph/9912039.



\bibitem{Du03QC}
W. D\"ur and H.-J. Briegel, Phys. Rev. Lett. {\bf 90}, 067901 (2003).

\bibitem{Ra04}
R. Raussendorf, S. Anders {\em et al.}, unpublished.

\bibitem{As99}
H. Aschauer and H.-J. Briegel, Phys. Rev. Lett. {\bf 88}, 047902 (2002); 
ibid Phys. Rev. A {\bf 66}, 032302 (2002).


\bibitem{As04}
H. Aschauer, PhD thesis, LMU Munich (2004).


\bibitem{Ja98}
D. Jaksch, H.-J. Briegel, J. I. Cirac, C. W. Gardiner and P. Zoller, 
Phys. Rev. Lett. {\bf 82}, 1975 (1999).

\bibitem{Gr02}
M. Greiner, O. Mandel, T. Esslinger, T.W. H\"ansch and I. Bloch, 
Nature {\bf 415}, 39 (2002).

\bibitem{Gr02b}
M. Greiner, O. Mandel, T. W. H\"ansch and I. Bloch, 
Nature {\bf 419}, 51 (2002).

\bibitem{Ma03}
O. Mandel, M. Greiner, A. Widera, T. Rom, T.W. H\"ansch, and I. Bloch, 
Phys. Rev. Lett. {\bf 91}, 010407 (2003).

\bibitem{Ma03b}
O. Mandel, M. Greiner, A. Widera, T. Rom, T. W. H\"ansch, and I. Bloch, 
Nature {\bf 425}, 937 (2003).

\bibitem{Ch04}
Kai Chen and Hoi-Kwong Lo, quant-ph/0404133.

\end{thebibliography}
\end{document}